**Modeling of High-Sensitivity SAW Magnetic Field Sensors with Au-$SiO_2$ Phononic Crystals**

*Mohsen Samadi*[1,*], *Jana Marie Meyer*[2], *Elizaveta Spetzler*[3], *Benjamin Spetzler*[4], *Jeffrey McCord*[3,5], *Fabian Lofink*[2,5,6], *and Martina Gerken*[1,5,*]

[1]Integrated Systems and Photonics, Department of Electrical and Information Engineering, Kiel University, Kaiserstraße 2, 24143 Kiel, Germany

[2]Fraunhofer Institute for Silicon Technology ISIT, 25524 Itzehoe, Germany

[3]Nanoscale Magnetic Materials - Magnetic Domains, Department of Materials Science, Kiel University, Kaiserstraße 2, 24143 Kiel, Germany

[4]Energy Materials and Devices, Department of Materials Science, Kiel University, Kaiserstraße 2, 24143 Kiel, Germany

[5]Kiel Nano, Surface and Interface Science (KiNSIS), Kiel University, Kaiserstraße 2, 24143 Kiel, Germany

[6]Microsystem Materials, Department of Materials Science, Kiel University, Kaiserstraße 2, 24143 Kiel, Germany

*E-Mail: mosa@tf.uni-kiel.de, mge@tf.uni-kiel.de

**Abstract:** The development of magnetic field sensors with high sensitivity is crucial for accurate detection of magnetic fields. In this context, we present a theoretical model of a highly sensitive surface acoustic wave (SAW) magnetic field sensor that utilizes phononic crystal (PnC) structures composed of Au pillars embedded within a $SiO_2$ guiding layer. We study rectangular and triangular PnCs and assess their potential for application in thin-film magnetic field sensors. In our design, the PnC is integrated into the $SiO_2$ guiding layer, preserving the continuous magnetostrictive layer and maximizing its interaction with the SAW. The sensor achieves nearly two orders of magnitude higher sensitivity compared to a continuous delay line of similar dimensions and an eight-fold improvement over the previous sensor design with PnCs composed of FeCoSiB pillars. The enhanced sensitivity is attributed to resonance effects within the PnC leading to an increased interaction between the SAW and the continuous FeCoSiB layer covering the PnC. Our results highlight the significant potential of incorporating PnCs into the guiding layer of SAWs for future high performance magnetic field sensors.

## 1. Introduction

Magnetic field sensors are essential tools for detection and accurate measurement of magnetic fields in various applications.[1–11] In particular, magnetic field sensors based on surface acoustic waves (SAW)[12–15] have been employed to detect minute changes in magnetic field through the $\Delta E$ effect, where the stiffness tensor of a magnetostrictive material changes in response to an external magnetic field.[16–20] This change induces modulation in the SAW phase, which is measured to determine the magnetic field strength.[21–25]

Unlike sensing approaches based on magnetoelectric cantilevers, which are limited by narrow bandwidths ranging from a few hertz to a few kHz,[26–28] SAW-based magnetic field sensors can detect small amplitude magnetic fields over a broader frequency range.[21–24] Recently, detection limits of 2.4 $nT/Hz^{1/2}$ at 10 Hz and 72 $pT/Hz^{1/2}$ at 10 kHz were experimentally achieved with a thin-film SAW magnetic field sensor, highlighting the potential of this configuration for highly sensitive magnetic field detection.[25] The sensor utilized a thin film of aluminum scandium nitride (AlScN) as the piezoelectric material on a silicon substrate. A $SiO_2$ layer was deposited onto the AlScN layer to confine acoustic waves near the surface, minimizing energy loss into the substrate and enhancing the interaction between the SAW and the magnetostrictive material. Additionally, the $SiO_2$ layer was mechanically polished to create a smooth surface for the deposition of the magnetostrictive FeCoSiB film, improving its soft magnetic properties.

Compared to bulk piezoelectric substrates such as quartz[22] and $LiNbO_3$[29], which are commonly used, thin-film SAW configurations offer reduced chip sizes and improved functionality, providing a more compatible solution for complementary metal-oxide semiconductor (CMOS) and micro-electro-mechanical system (MEMS) technologies.[30] Thin-film Aluminum Nitride (AlN) is recognized as a highly promising piezoelectric material due to its high acoustic wave velocity, thermal stability, and CMOS compatibility enabled by low-temperature deposition.[31–35] Alloying AlN with transition metals such as Scandium (Sc) improves its electromechanical coupling, making it particularly suitable for high-frequency applications.[36,37]

SAWs can drive magnetization dynamics in ferromagnetic materials through spin-phonon coupling, a process enabled by SAW-driven ferromagnetic resonance (FMR).[38–41] This phenomenon allows for control over both spin and SAW dynamics, giving rise to effects such as nonreciprocity of SAWs.[42–46] Periodic structures further provide precise control over wave propagation and dispersion. For instance, photonic crystals (PhCs) manipulate electromagnetic waves to create peculiar phenomena such as photonic bandgaps[47–49], resonant field enhancement and localization[50–52], and slow-light effects[53–55], which have been leveraged to achieve high sensitivities in PhC-based sensors.[56–58] Similarly, phononic[59] and magnonic[60] crystals enable the control of acoustic and spin waves, respectively. The ability to manipulate waves has made these periodic structures a powerful tool for improving sensor sensitivity by facilitating strong wave localization and enhancing wave-material interactions.

Phononic crystals (PnCs) are periodic structures composed of two materials with different acoustic impedance values.[59,61–64] By tailoring the geometry and material composition of PnCs, precise control over the propagation and dispersion of acoustic waves can be achieved.[65,66] A prior study[67] demonstrated that integrating carefully designed PnCs into SAW magnetic field sensors significantly enhances their sensitivity to magnetic field variations. The PnC structure consisted of a 2D square array of FeCoSiB pillars deposited onto a $SiO_2$ guiding layer. Nevertheless, patterning the magnetostrictive layer results in a considerable material loss, thereby reducing the interaction between the SAW and the magnetostrictive material.

In this groundbreaking study, adopting a configuration similar to that proposed in our previous research[25], we leverage the remarkable piezoelectric properties of AlScN thin films with an integrated PnC structure to present a design and theoretical model for a thin-film SAW magnetic field sensor with a drastically improved sensitivity. The PnC consists of a 2D lattice of Au pillars embedded within a $SiO_2$ guiding layer, topped with a continuous layer of magnetostrictive material. The resonant SAWs propagate through the PnC, confined to the top surface of the guiding layer that surrounds the PnC, thereby strongly enhancing the interaction with the unpatterned magnetostrictive layer.

We evaluate the changes in the stiffness tensor of the magnetostrictive layer in response to small magnetic field variations using a magnetoelastic macrospin model. The resulting data is then used as input for the electromechanical model to estimate changes in the phase velocity of the PnC modes and assess the sensor's sensitivity. Through precise alignment of the PnC resonant modes with the frequency of the input SAW, we demonstrate that a small change in the applied magnetic field results in a substantial change in the phase velocity, yielding remarkably high sensitivity.

The paper is structured as follows. Section 2 introduces the novel sensor design, explaining the thin-film configuration and the PnC structure. Section 3 describes the theoretical model, focusing on the electromechanical and magnetoelastic models as well as the sensor's sensitivity. Section 4 presents the results, divided into subsections that examine the band structures of the PnCs, the displacement of SAWs through the PnCs, and the use of the ΔE effect as a mechanism for magnetic field detection. In section 5 the results are discussed in a bigger picture. Finally, the improvements in comparison to existing structures are highlighted in section 6.

## 2. Sensor Design

Figures 1(a) and 1(b) schematically display the 3D and cross-sectional views of the sensor. A thick silicon substrate is coated by a 1 µm-thick AlScN piezoelectric layer. Two interdigital transducers (IDT) are devised on the AlScN layer to generate and detect a Rayleigh wave at a center frequency of $f = 250$ MHz. A $SiO_2$ layer is deposited onto the piezoelectric layer acting as a guiding medium for the SAW. A rectangular or triangular lattice of Au pillars is embedded within the $SiO_2$ guiding layer between the two IDTs, as illustrated in Figure 1(c) and 1(d). The pillars are of the same thickness as the guiding layer ($h_p = h_{GL} = 4.5$ µm). The entire PnC structure is then covered by a continuous magnetostrictive FeCoSiB layer with a thickness of $h_{MS} = 200$ nm.

The top views of the rectangular and triangular PnCs are depicted in Figure 1(c) and 1(d). Rectangular supercells, marked by dashed boxes, were selected in both configurations to calculate the PnC band structures. This approach enables the construction of the complete structures by repeating the supercells along the $x$- and $y$-axes. To ensure a valid comparison, the primary lattice constant $a$, defined as the distance between the nearest-neighbor pillars along the $x$-axis, is identical in both PnCs. The secondary lattice constant $b$ along the $y$-axis was defined as $b = 2aq$ for the rectangular PnC and $b = \sqrt{3}aq$ for the triangular PnC. Here, the tuning coefficient $q$ adjusts the lateral spacing between pillars relative to their longitudinal distance to maximize sensitivity.

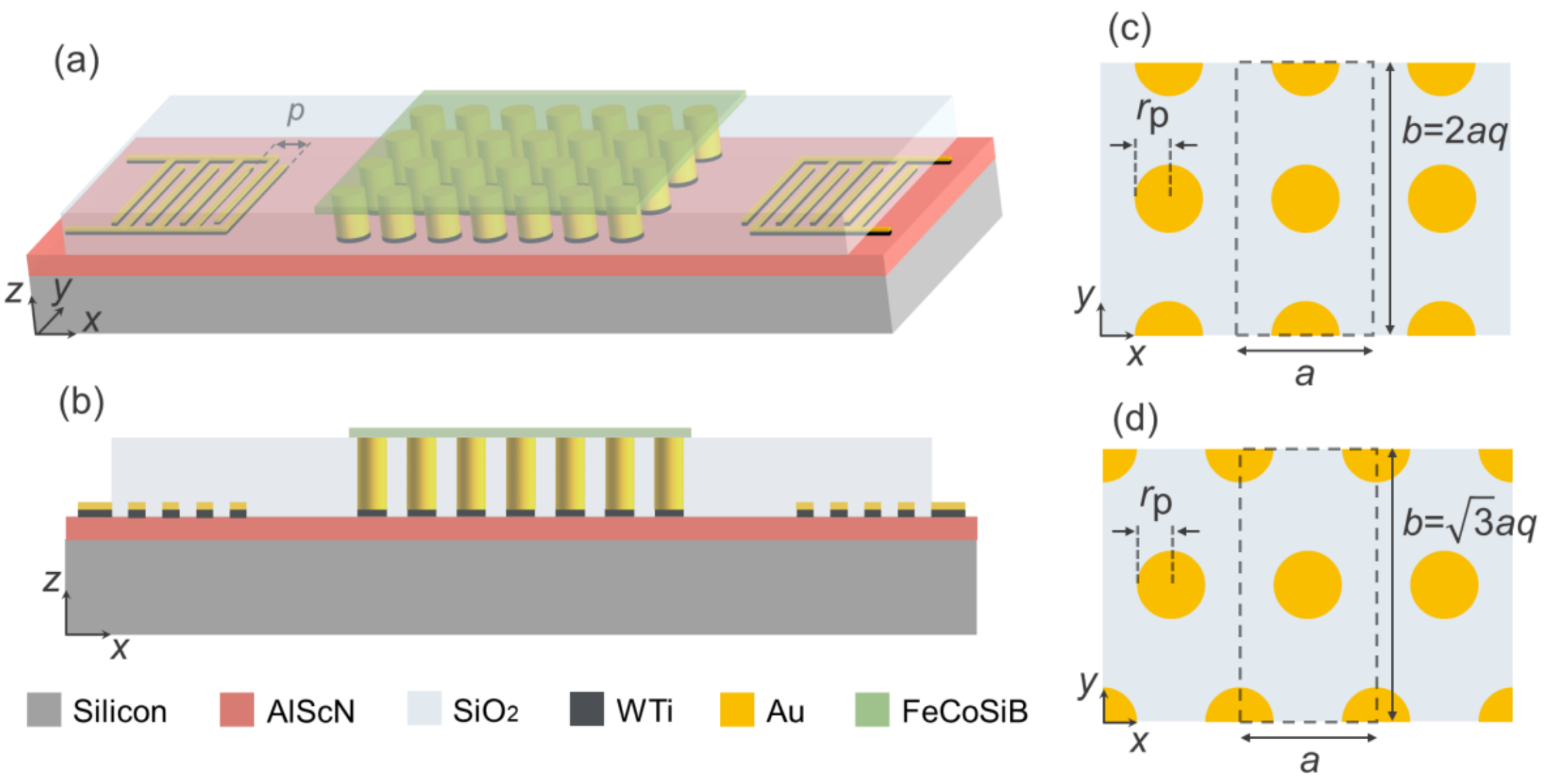

**Figure 1.** (a) 3D and (b) cross-sectional view of the SAW magnetic field sensor with Au-$SiO_2$ PnC. A rectangular or triangular lattice of Au pillars is patterned within a $SiO_2$ guiding layer to build a PnC between the IDTs. Top views of the (c) rectangular and (d) triangular PnCs. Rectangular super cells of the two PnCs are indicated by dashed lines. $a$ and $b$ denote the lattice constants of the PnCs along the $x$- and $y$-axis, while $r_p$ represents the radius of the pillars.

The main focus of this study is the design and theoretical modeling of a SAW magnetic field sensor with integrated PnCs to enhance sensitivity. However, a feasible fabrication process has also been outlined based on the design parameters obtained from simulations and is currently in progress. The fabrication process begins with the sputtering of an AlScN layer onto a silicon wafer. For the IDTs and the plating base, WTi/Au is sputtered and structured using wet chemical etching. The $SiO_2$ guiding layer is then deposited using plasma-enhanced chemical vapor deposition (PECVD) and patterned to create holes matching the dimensions of the designed PnC pillars. Next, an electroplating process is used to fill the structured holes with Au, forming the PnC pillars. A thin passivation layer of $SiO_2$ is then deposited using PECVD and smoothed via chemical mechanical polishing. Finally, the FeCoSiB magnetostrictive layer is sputtered onto the PnC, completing the fabrication process.

## 3. Theoretical Model

### 3.1. Equation system

The sensor's electromechanical response is characterized by solving a set of coupled differential equations through the finite element method (FEM). The equation of motion relates the displacement vector $\mathbf{u}$ to the divergence of the mechanical stress tensor $\boldsymbol{\sigma}$:[68,69]

$$\nabla.\boldsymbol{\sigma} = -\rho\omega^2\mathbf{u}. \quad (1)$$

Here, $\rho$ and $\omega$ are the mass density and the angular frequency, respectively. Assuming that free charges and eddy currents are negligible, Maxwell's electrostatic equation is expressed as[70]

$$\nabla.\mathbf{D} = 0, \quad (2)$$

where **D** represents the electric flux density vector. The electric field vector **E** is derived from the gradient of the electric potential *V*:

$$\mathbf{E} = -\nabla V. \quad (3)$$

The mechanical equation of motion (Equation (1)) is coupled to the electrostatic equation (Equation (2)) through the constitutive piezoelectric equations in the stress-charge form:[70,71]

$$\boldsymbol{\sigma} = \mathbf{C}\boldsymbol{\varepsilon} - \mathbf{e}^{\mathrm{T}}\mathbf{E}, \quad (4)$$

$$\mathbf{D} = \mathbf{e}\boldsymbol{\varepsilon} - \boldsymbol{\varepsilon}_{\mathrm{el}}\mathbf{E}. \quad (5)$$

In these equations, $\boldsymbol{\varepsilon}$ is the strain tensor, $\mathbf{C}$ is the mechanical stiffness tensor, while $\mathbf{e}$ and $\boldsymbol{\varepsilon}_{\mathrm{el}}$ denote the piezoelectric coupling and electrical permittivity tensors, respectively.

### 3.2 Modeling approach

To investigate the $\Delta E$ effect, we employed a semi-analytical magnetoelastic macrospin model to estimate the magnetic field-dependent stiffness tensor of FeCoSiB. Details of this approach are provided in another study.[20] The resulting magnetic field-dependent stiffness tensor was then used in Equation 4, as input to the electromechanical model developed in COMSOL Multiphysics®[72] to simulate the sensor's response. The material parameters used in the magnetoelastic model are given in Appendix A.

We employed COMSOL Multiphysics® to calculate the band structures of the PnCs, limiting our analysis to a single supercell, as indicated by the dashed rectangles in Figure 1(c) and 1(d), to save time and computational resources. Bloch-Floquet boundary conditions were applied at the supercell boundaries to simulate infinite periodicity in both *x* and *y* directions.[64,73] By varying the wavevector along the Γ-X direction of the first Brillouin zone and solving the eigenvalue problem, we obtained the eigenfrequencies for each wavevector. These eigenfrequencies indicate the acoustic modes permitted to propagate through the PnCs.

Moreover, we modeled PnCs with a finite number of periodicities ($n = 20$) integrated into the delay line of a SAW magnetic field sensor. We applied an alternating voltage to an IDT to generate a Rayleigh wave with a specific center frequency. The PnC parameters were carefully designed to align a PnC mode with the Rayleigh wave generated by the IDT. We calculated the average displacement across the top surface of the PnCs, which was coated with an FeCoSiB layer, at different frequencies. To model infinite periodic repetitions and sufficiently long IDT fingers, we applied a periodic boundary condition along the $y$-axis. Additionally, perfectly matched layers (PML) were imposed in both the $x$ and $z$ directions to prevent unwanted reflections from the domain boundaries.

### 3.3 Sensor's sensitivity

Magnetic sensitivity describes the variation in the stiffness tensor of the magnetostrictive layer in response to changes in the applied magnetic field magnitude. At a specific bias magnetic field magnitude $|H| = H_0$, a small field variation $\Delta H$ induces a change in the stiffness tensor $\mathbf{C}$ of the magnetostrictive material. Since different components of the stiffness tensor $\mathbf{C}$ exhibit distinct responses to the applied magnetic field, magnetic sensitivity is defined by a tensor $\mathbf{S}_{\mathrm{mag}}$ of the same order and dimensions as $\mathbf{C}$. Each component of the magnetic sensitivity tensor represents the rate of change of its corresponding stiffness tensor component with respect to the magnetic field magnitude $|H|$, expressed as $S_{\mathrm{mag},ij} = \partial \mathrm{C}_{ij}/\partial |H|$.

Structural sensitivity quantifies how changes in the stiffness tensor $\mathbf{C}$ induced by the magnetic field affect the SAW velocity $v$. It is strongly influenced by the design of the delay line through which the wave propagates. For instance, increasing the thickness of the guiding layer beneath the magnetostrictive material in a delay line configuration has been shown to enhance the structural sensitivity.[21] Additionally, a recent study demonstrated that the structural sensitivity of a SAW magnetic field sensor can be improved by replacing the continuous magnetostrictive layer with a periodic array of magnetostrictive pillars, forming a PnC.[67] The SAW velocity $v$ is influenced differently by each component of the stiffness tensor $\mathbf{C}$. Therefore, structural sensitivity is expressed as a tensor $\mathbf{S}_{\mathrm{str}}$ of the same order and dimensions as $\mathbf{C}$, with each component $S_{\mathrm{str},ij} = \partial v/\partial \mathrm{C}_{ij}$ representing the derivative of the SAW velocity with respect to the corresponding components of the stiffness tensor.

The combined effect of the magnetic and structural sensitivity is represented by a scalar value, referred to as the magneto-structural sensitivity, $S_{\mathrm{mag-str}}$. This scalar quantifies how the SAW velocity varies in response to changes in the magnetic field magnitude. It is calculated by summing the products of the corresponding components of the two sensitivity tensors, $\mathbf{S}_{\mathrm{mag}}$ and $\mathbf{S}_{\mathrm{str}}$, over all indices:

$$S_{\mathrm{mag-str}} = \frac{\partial v}{\partial |H|} = \sum_{i,j} S_{\mathrm{mag},ij} \cdot S_{\mathrm{str},ij}, \qquad (6)$$

where $S_{\mathrm{mag},ij}$ and $S_{\mathrm{str},ij}$ denote the components of the magnetic and structural sensitivity tensors and the summation is performed over all indices $i$ and $j$.

The phase $\varphi$ of a SAW with a frequency $f$ and a velocity $v$, detected at the end of a delay line of length $l$, is given by:

$$\varphi = 2\pi f \frac{l}{v}. \tag{7}$$

At a fixed frequency, any change in the SAW velocity induces a phase delay in the output signal. The extent of this phase delay per unit change in the SAW velocity, referred to as the geometric sensitivity $S_{\text{geo}}$, is expressed as:

$$S_{\text{geo}} = \frac{\partial \varphi}{\partial v} = -2\pi f \frac{l}{v^2}. \tag{8}$$

The overall sensitivity of the sensor is determined by the product of the magneto-structural and geometric sensitivity, quantifying the sensor's output phase delay in response to changes in the magnetic field magnitude:

$$S = \frac{\partial \varphi}{\partial |H|} = \frac{\partial v}{\partial |H|} \cdot \frac{\partial \varphi}{\partial v} = S_{\text{mag-str}} \cdot S_{\text{geo}}. \tag{9}$$

## 4. Results

### 4.1. Band structures of the PnCs

In this section, we calculate the band structures and analyze the dispersion characteristics of PnCs with either a rectangular or triangular arrangement of Au pillars embedded within a $SiO_2$ guiding layer, as depicted in Figure 1(c) and 1(d). Band diagrams of the rectangular and triangular PnCs with $a$ = 6 µm, $r_p$ = 1.5 µm, calculated along the Γ-X direction of the first Brillouin zone, are shown in Figure 2 and S1, respectively. The gray-shaded area in the band diagrams represents the sound cone. Acoustic modes that lie within this cone radiate into the bulk material, whereas surface acoustic modes appear only outside the sound cone. The tuning coefficient is varied between $q$=1 and $q$=2 by a step size of 0.2, to investigate the effect of lateral distance between pillars with respect to their longitudinal distance.

To provide a clear understanding of each mode's characteristics, Figure 3 presents the band diagram of a rectangular PnC with a fixed tuning coefficient of $q$=2, with each mode labeled by a number. The displacement profile of each mode is displayed within a single supercell at a wavevector $k$ near the edge of the first Brillouin zone. The band diagram and modal displacement profiles for a triangular PnC with a tuning coefficient of $q$=2 are also presented in Figure S2. Based on the displacement profiles, lower-order modes exhibit mostly uniaxial oscillations, while higher-order modes induce combined longitudinal and vertical surface displacements. Modes 1 and 2 (blue curves) oscillate vertically along the $z$-axis, whereas modes 3 and 4 (green curves) generate shear-horizontal displacements along the *y*-axis, representing Love modes. Modes 5 and 6 (red curves) primarily induce longitudinal displacements along the *x*-axis. In contrast, modes 7 and 8 (magenta curves) exhibit both longitudinal and vertical oscillations in the *x*-*z* plane, corresponding to Rayleigh modes. To enable SAW propagation through the PnC, it is essential that the selected PnC mode matches in both frequency and polarization with the Rayleigh wave launched by the IDT. In the current PnC configuration, Rayleigh waves are utilized due to their out-of-plane oscillations, which enable efficient coupling with the continuous FeCoSiB layer on top of the $SiO_2$ guiding layer, resulting in higher sensitivity. While Love waves could be explored as an alternative, their purely in-plane displacements pose challenges for efficient coupling with the top FeCoSiB layer in this specific design. Accordingly, the following sections of the paper focus on mode 7, a Rayleigh mode with a center frequency of approximately 250 MHz.

Integrating PnCs into a SAW magnetic field sensor can significantly enhance its sensitivity. To demonstrate this, we compared the band diagram of a rectangular PnC with a tuning coefficient of $q$=2 to that of a continuous guiding layer with similar dimensions, as shown in Figure S3. The band diagram of the continuous guiding layer was calculated using a supercell identical in dimensions to that in Figure 1(b), but without Au pillars. The band diagram of the PnC exhibits nearly flat bands with extremely low group velocities near the edge of the first Brillouin zone (X point), in contrast to the continuous guiding layer, where surface acoustic modes appear as linear bands with a fixed group velocity for all wavevectors along the Γ-X direction. The slow-wave phenomenon in PnCs arises from the resonant coupling of local modes between neighboring pillars at specific frequencies, forming collective guided surface modes with reduced group velocity. This leads to an effectively longer interaction length between the SAW propagating through the PnC and the continuous magnetostrictive layer covering the guiding layer's top surface. By fine-tuning the PnC design parameters, the edge of a PnC mode can be precisely aligned with the frequency of the input SAW generated by an IDT. Consequently, small changes in the magnetic field induce larger variations in the SAW phase velocity, significantly enhancing the sensor's sensitivity.

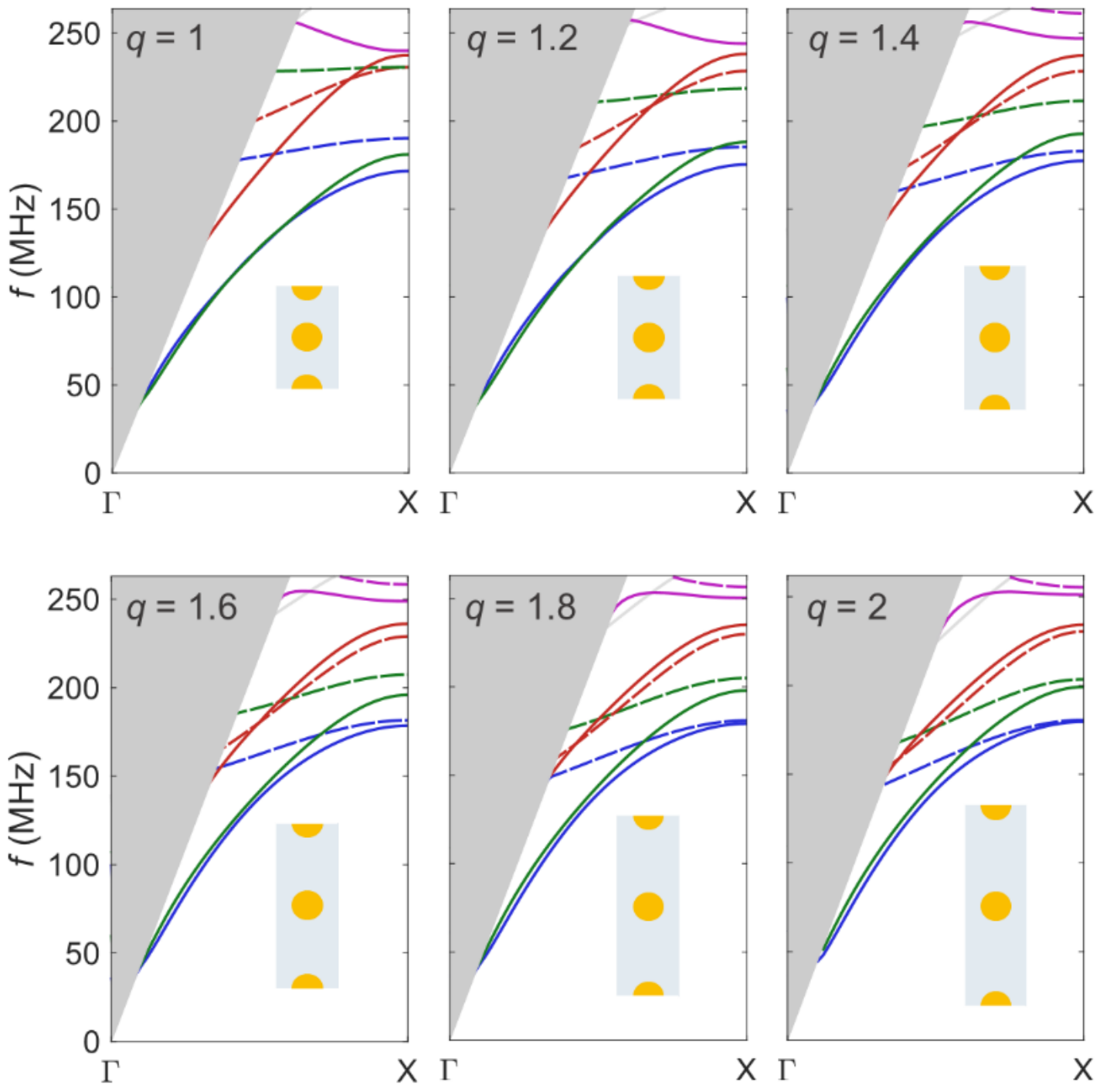

**Figure 2.** Band structures of the rectangular PnC, as illustrated in Figure 1(b), with $a$ = 6 μm, $r_p$ = 1.5 μm, and different values of the tuning coefficient: $q$=1, $q$=1.2, $q$=1.4, $q$=1.6, $q$=1.8 and $q$=2. The lattice constant along the $y$-axis is expressed as $b = 2aq$ for the rectangular PnC. All diagrams were calculated along the Γ-X direction of the first Brillouin zone. The grey-shaded area outlines the sound cone. Insets display the top views of the rectangular supercells used to calculate the band structures.

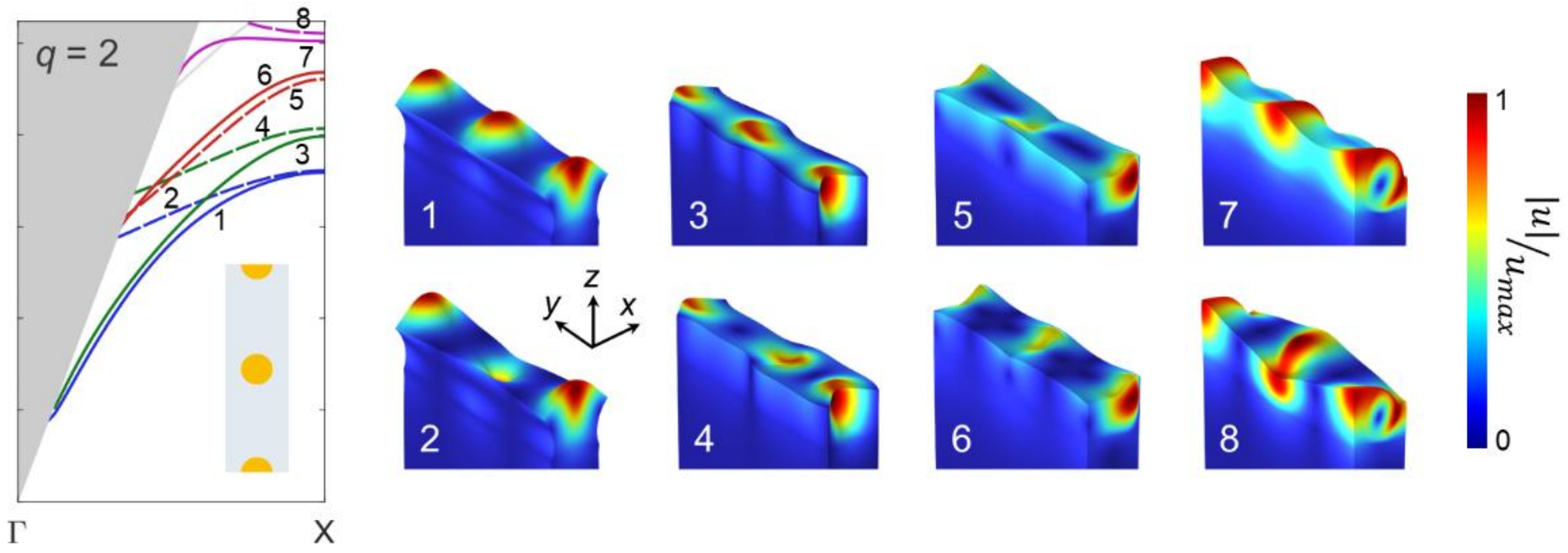

**Figure 3.** Band structure of the rectangular PnC with $a$ = 6 μm, $r_p$ = 1.5 μm, and $q$=2, calculated along the Γ-X direction of the first Brillouin zone. Distinct surface acoustic modes are labeled by numbers, with corresponding normalized modal displacements ($|u|/u_{max}$) displayed within a single supercell at a wavevector $k$ near the edge of the first Brillouin zone. Lower-order modes exhibit almost uniaxial oscillations: modes 1 and 2 (blue) along the $z$-axis, modes 3 and 4 (green) along the $y$-axis, and modes 5 and 6 (red) along the $x$-axis. In contrast, higher-order modes 7 and 8 (magenta) exhibit coupled longitudinal and vertical oscillations in the $x$-$z$ plane.

In both rectangular and triangular configurations, an increase in the tuning coefficient $q$, which corresponds to a larger lateral distance between the pillars relative to the longitudinal lattice constant $a$, reduces the slopes of all bands, including mode 7. To illustrate this effect, Figures 4(a) and S4(a) present the dispersion curves of mode 7 near the edge of the first Brillouin zone for the rectangular and triangular PnCs, respectively. These curves are plotted for different values of the tuning coefficients: $q$=1, $q$=1.2, $q$=1.4, $q$=1.6, $q$=1.8 and $q$=2. From these dispersion curves, the group velocity ($v_g = 2\pi \frac{\partial f}{\partial k}$) of mode 7 was calculated at a certain wavenumber $k = 0.8\pi/a$ for each value of $q$. Figures 4(b) and S4(b) show the group velocity variations with $q$ for the rectangular and triangular PnCs, respectively. In both configurations, a noticeable decrease in group velocity is observed as $q$ increases. At $q$=2, the group velocity reaches approximately 100 m/s for the rectangular PnC and 150 m/s for the triangular PnC.

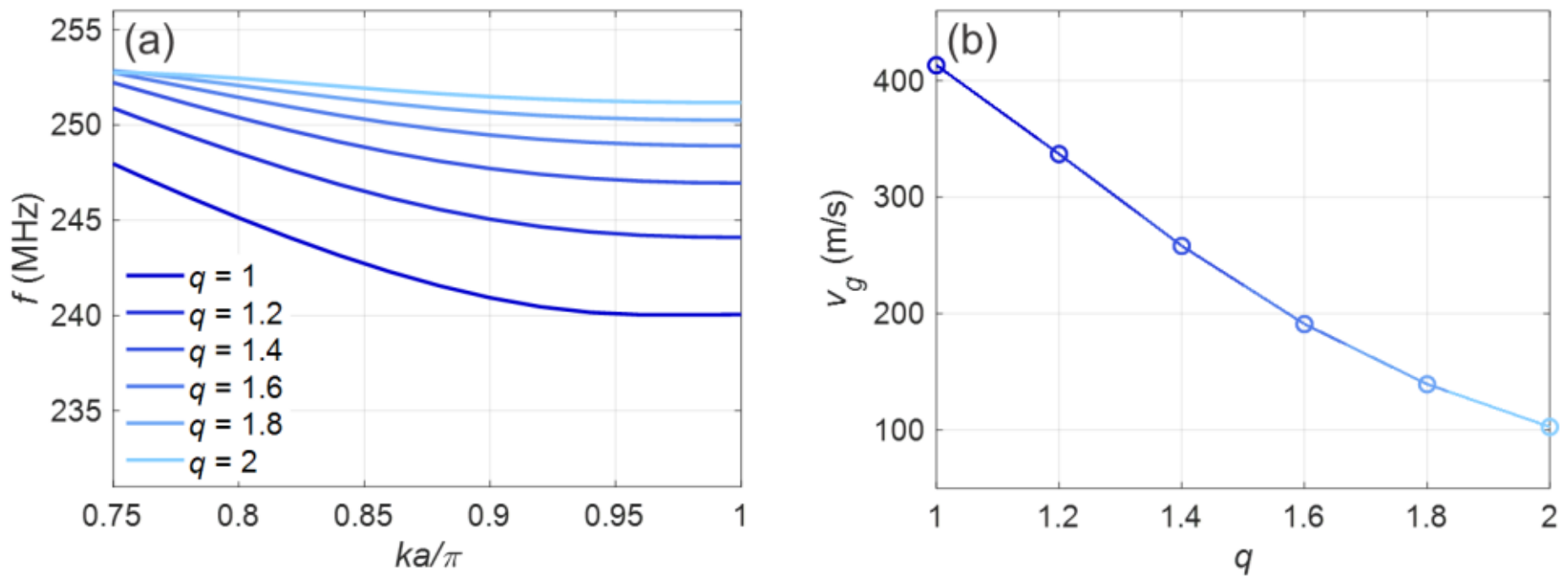

**Figure 4.** (a) Dispersion curves of mode 7 near the edge of the first Brillouin zone for the rectangular PnC at different values of the tuning coefficients: $q$=1, $q$=1.2, $q$=1.4, $q$=1.6, $q$=1.8 and $q$=2. (b) Group velocity variations with $q$ for the rectangular PnC, calculated from the dispersion curves at $k = 0.8\pi/a$.

### 4.2. Displacement across the PnCs

To model the propagation of SAWs through the PnCs, either a rectangular or triangular PnC with $a$ = 6 μm, $r_p$ = 1.5 μm, $q$ = 2, and finite number of periodicities ($n$ = 20) along the $x$-axis was integrated into the delay line of a SAW magnetic field sensor, as shown in Figure 5(b) and S5(b). A spatially alternating electric potential was applied to an IDT composed of 12 pairs of Au split-finger structures with a periodicity of 16 μm, a finger width of 4 μm, and a thickness of 150 nm to generate a Rayleigh wave with a center frequency of 250 MHz. The average displacement across the top surface of the PnCs, coated with an FeCoSiB layer, was then calculated at different frequencies. Without the PnCs, the average displacement spectrum features a broad, barely-visible peak centered around 250 MHz (red curve in Figure 5(a) and S5(a)). However, when a PnC is integrated into the delay line, the average displacement spectra reveal sharp peaks, as shown by the blue curve in Figure 5(a) and the green curve in Figure S5(a). These sharp peaks at certain frequencies are attributed to resonance effects induced by the periodic arrangement of pillars within the PnCs. Maximum values of approximately 0.11 nm and 0.05 nm, per unit of the electric potential applied to the IDT, are achieved at the resonance frequencies of the rectangular and triangular PnCs, respectively. Figure 5(b) and S5(b) depict the spatial displacement along each type of PnC at their respective resonance frequencies, showing that both PnCs allow Rayleigh waves to be resonantly guided, thereby amplifying oscillations across their top surfaces. At other frequencies, however, PnCs block the incoming Rayleigh wave, leading to minimal oscillations along the delay line.

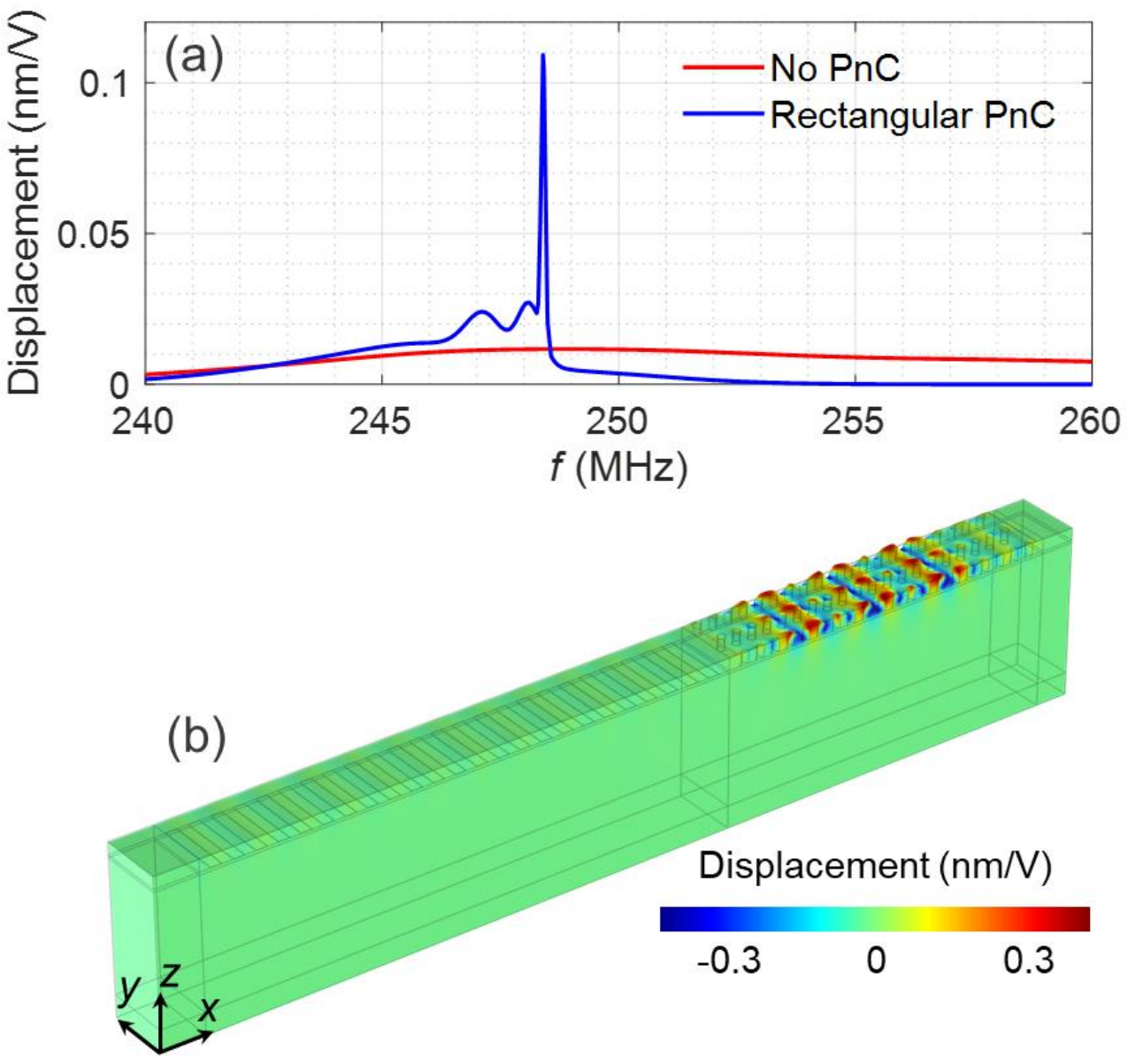

**Figure 5.** (a) Average displacement spectra across the top surface of the delay line with the rectangular PnC with $a$ = 6 µm, $r_p$ = 1.5 µm, and $q$=2 (blue curve). The average displacement spectra for a continuous delay line without PnC is also plotted for comparison (red curve). (b) The spatial displacement along the rectangular PnC at its resonance frequency. The displacement values are normalized to the amplitude of the AC electric potential applied to the IDT. The PnC has 20 periodicities along the $x$-axis ($n$ = 20) and the IDT comprises 12 pairs of Au split-finger structures with a periodicity of 16 µm, a finger width of 4 µm, and a thickness of 150 nm.

### 4.3. Δ*E* effect

We used the magnetoelastic macrospin model to calculate the stiffness tensor components of the FeCoSiB layer under various magnetic field strengths. We assumed that the applied magnetic field was oriented along the *y*-axis, aligned with the hard axis of the FeCoSiB layer. Figure 6 presents the primary non-zero magnetic field-dependent components of the FeCoSiB stiffness tensor, i.e., $C_{11}$, $C_{12}$, $C_{16}$ and $C_{66}$, along with the derivatives of each component with respect to the applied magnetic field. Other non-zero components are excluded, as they either depend on and can be derived from these four components ($C_{22}$ = $C_{11}$, $C_{21}$ = $C_{12}$, $C_{61}$ = $C_{16}$, $C_{62}$ = $C_{26}$ = $-C_{16}$) or remain invariant under an in-plane magnetic field ($C_{13}$ = $C_{31}$ = $C_{23}$ = $C_{32}$ = 86.54 GPa, $C_{33}$ = 201.92 GPa and $C_{44}$ = $C_{55}$ = 57.69 GPa).[20] These four components, along with their dependent components, exhibit significant variations with the applied magnetic field. Among them, $C_{16}$ is zero in the absence of an external magnetic field (at $|H| = 0$). The other three components, $C_{11}$, $C_{12}$, and $C_{66}$, have non-zero baseline values at zero magnetic field, to which a field-dependent term, varying with magnetic field strength, is added.

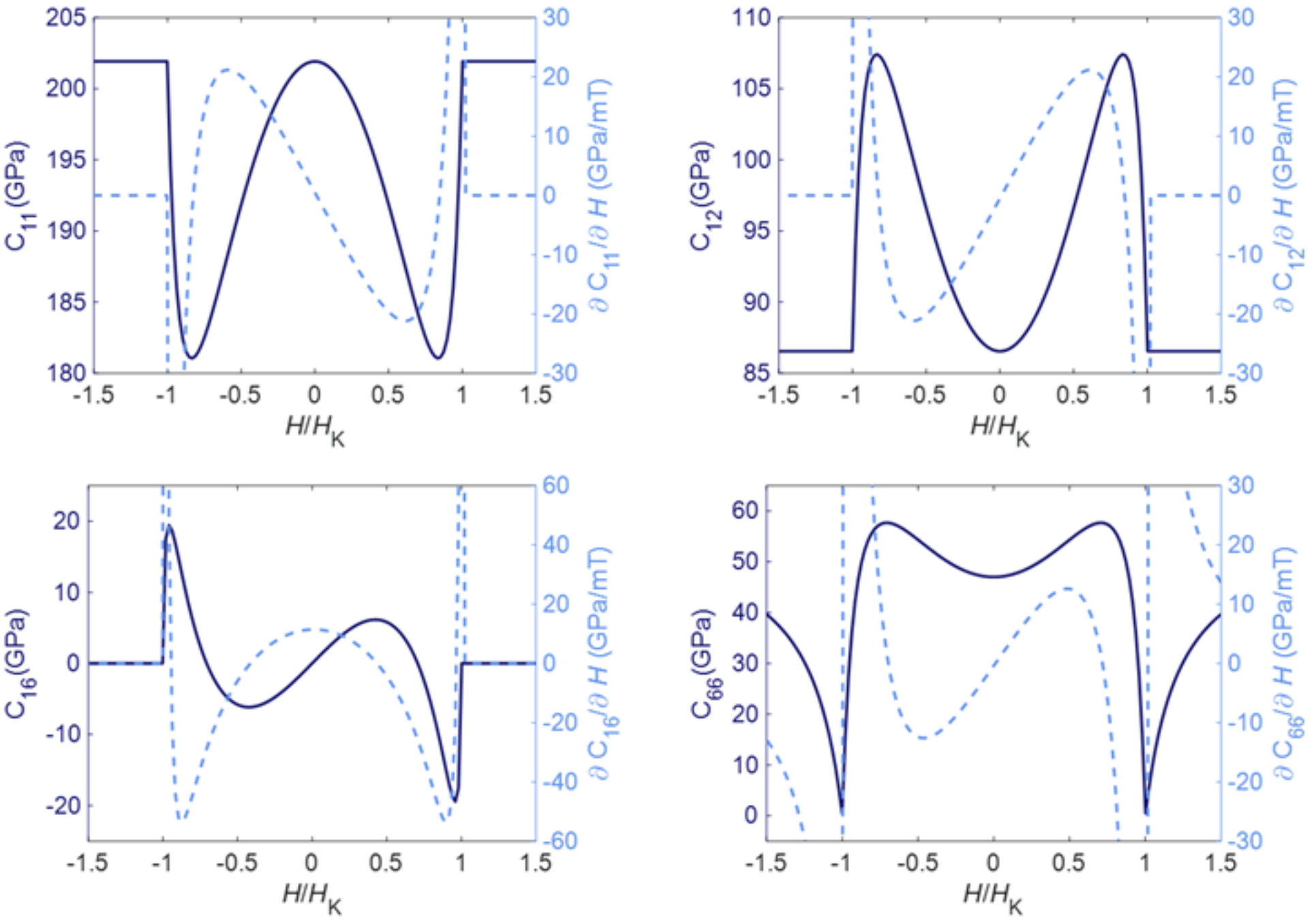

**Figure 6.** Non-zero magnetic field-dependent components of the FeCoSiB stiffness tensor, i.e., $C_{11}$, $C_{12}$, $C_{16}$ and $C_{66}$ (solid dark blue curves), and their derivatives with respect to the applied magnetic field magnitude, $\partial C_{ij}/\partial |H|$ (dashed light blue curves).

The magnetic sensitivity of each stiffness tensor component is highest when the derivative of that component with respect to the applied magnetic field magnitude, $\partial C_{ij}/\partial |H|$, is maximized. For our analysis, we selected a magnetic field magnitude of $H_0 = 0.7 H_K \sim 1.65$ mT, where $\partial C_{11}/\partial |H|$ and $\partial C_{12}/\partial |H|$ reach their maximum values and $C_{16}$ is sufficiently small to be neglected. Additionally, $C_{66}$ is maximized at this field value, while $\partial C_{66}/\partial |H|$ remains negligible. This implies that small changes in the magnetic field around this specific value primarily affect the longitudinal and transverse deformations, with negligible effects on the shear deformations. The values of the primary stiffness tensor components, $C_{11}$, $C_{12}$, $C_{16}$ and $C_{66}$, and their corresponding magnetic sensitivity ($\partial C_{ij}/\partial|H|$) at the magnetic field of $H_0$ are given in Table 1.

**Table 1.** Calculated values for the primary stiffness tensor components ($C_{ij}$) and their corresponding magnetic sensitivity ($\partial C_{ij}/\partial|H|$) at the bias magnetic field of $H_0 = 0.7H_K$.

| | $C_{11}$ | $C_{12}$ | $C_{16}$ | $C_{66}$ |
|---|---|---|---|---|
| $C_{ij}$ (GPa) | 184 | 104.4 | 0.2 | 57.7 |
| $\frac{\partial C_{ij}}{\partial\|H\|}$ (GPa/mT) | -19.5 | 19.5 | -25.3 | 2 |

We applied a small magnetic perturbation of $\Delta H = 100$ μT over the bias magnetic field of $H_0$ and estimated the stiffness tensor components for the magnetic field values $H^+ = H_0 + \Delta H/2$ and $H^- = H_0 - \Delta H/2$. We subsequently computed the band structure of the rectangular PnC for the stiffness tensors corresponding to these two field values, given as $\mathbf{C}^+$ and $\mathbf{C}^-$ in Appendix A. Figure 7 presents the resulting dispersion curves for the PnC mode $m_7$, plotted near the edge of the first Brillouin zone. The dispersion curves for the other PnC modes ($m_1$ to $m_8$) are also computed and displayed in Figure S6. From these dispersion curves, the magneto-structural sensitivity $S_{\text{mag-str}}$ of each mode was estimated as the phase velocity change $\Delta v$ in response to a $\Delta H = 100$ μT variation in the magnetic field. The results are presented as $S_{\text{mag-str}}\left(\frac{\text{m/s}}{\text{mT}}\right)$ in Table 2. Generally, the higher-order modes exhibit larger magneto-structural sensitivities due to their flatter profiles, and therefore smaller group velocities, comparing to the lower-order modes. In particular, mode $m_7$ demonstrates the highest sensitivity to magnetic field variations, with a value of 1588 $\left(\frac{\text{m/s}}{\text{mT}}\right)$, which is 66 times higher than the sensitivity obtained for a continuous guiding layer without the PnC.

Furthermore, we calculated the resonance frequency of mode $m_7$ at different magnetic field values around $H_0$ to analyze the frequency shift of this particular mode under varying magnetic perturbation strengths. As depicted in Figure S7, for perturbation strengths in the range of a few hundred μT, the resonance frequency scales linearly and inversely with the applied magnetic field, with a change rate of approximately 1 MHz/mT, shifting to lower frequencies at higher magnetic field values.

For a rectangular PnC with 20 pillars along the *x*-axis, resulting in a total length of $l$ = 120 µm, as illustrated in Figure 5, our calculations yield a geometric sensitivity of $S_{\text{geo}}$~ -1.19 °s/m at the frequency $f$ = 250 MHz and the phase velocity $v$ ~ 3017 m/s for mode $m_7$. For this mode, an overall sensitivity of $S$ ~ 1884 °/mT can be achieved, which is nearly two orders of magnitude greater than that of a continuous delay line of the same dimensions, and over 8 times higher than the sensitivity limits reached in our previous study using PnC structures composed of FeCoSiB pillars[67], normalized to the delay line length of the present study.

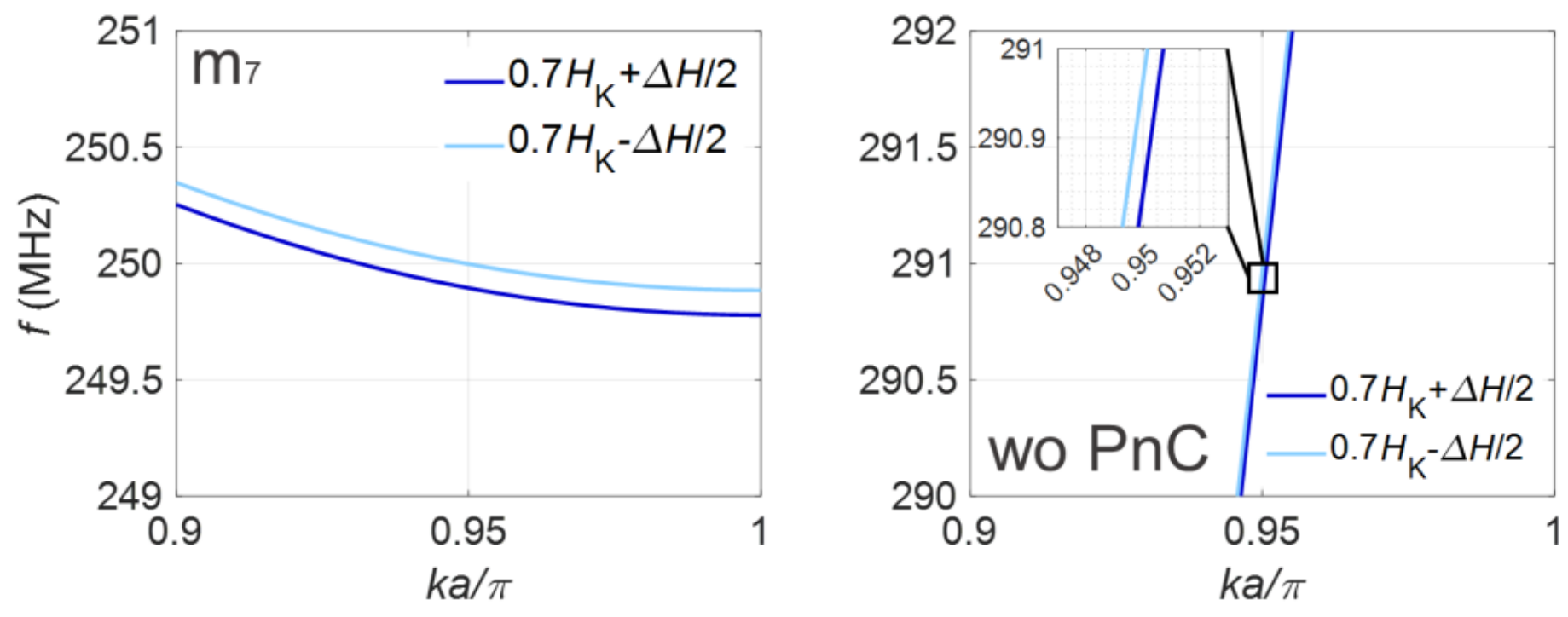

**Figure 7.** Dispersion curves of the rectangular PnC mode $m_7$ displayed near the edge of the first Brillouin zone, at $H^+ = 0.7H_K + \Delta H/2$ and $H^- = 0.7H_K - \Delta H/2$, with $\Delta H$ = 100 µT. The dispersion curves of the lowest-order mode of a continuous delay line without PnCs are also provided for comparison.

**Table 2.** Phase velocity change of the rectangular PnC modes $m_1$ to $m_8$ per unit change of the applied magnetic field, $S_{\text{mag-str}}\left(\frac{\text{m/s}}{\text{mT}}\right)$, derived from the data in Figure S6.

| | $m_1$ | $m_2$ | $m_3$ | $m_4$ | $m_5$ | $m_6$ | $m_7$ | $m_8$ | wo PnC |
|---|---|---|---|---|---|---|---|---|---|
| $S_{\text{mag-str}}\left(\frac{\text{m/s}}{\text{mT}}\right)$ | 219 | 220 | 365 | 625 | 563 | 720 | 1588 | 950 | 23 |

A comprehensive study of all geometric parameters, including the total length of the PnC, pillar diameter, and thickness, and their impact on sensor sensitivity is beyond the scope of this work. Here, we specifically investigate the effect of PnC thickness on sensor sensitivity. The results show that the magneto-structural sensitivity of modes $m_1$ to $m_6$ decreases as thickness increases. This reduction is partly due to the fact that the sensitivity of PnC modes scales with their resonance frequency, which shifts to lower values as thickness increases. Normalizing the sensitivity of each mode to its resonance frequency reveals that while modes $m_1$ and $m_2$ remain unaffected by thickness variations, modes $m_3$ to $m_6$ exhibit reduced normalized sensitivity for thicker PnCs due to the weaker interaction of their predominantly in-plane oscillations with the top magnetostrictive layer. Higher-order modes exhibit a different trend, with mode $m_7$ reaching its maximum sensitivity at $h = 0.5a$ (3 µm), suggesting an optimal thickness for maximizing sensor performance (Table S1).

## 5. Discussion

The band diagrams illustrate how adjusting the lateral distance between pillars relative to the longitudinal distance, controlled by the coefficient $q$, affects the band characteristics, such as slope and frequency, in both rectangular and triangular PnCs. However, these two types of PnCs exhibit different responses to changes in $q$ due to their distinct lattice symmetries. In the rectangular PnC, increasing the lateral spacing reduces interactions between the pillars along the $y$-axis. For sufficiently large values of $q$ (i.e., $q$>2), this configuration approximates a quasi-1D arrangement of pillars, resulting in degenerate modes at the X-point. As $q$ is reduced, the coupling between the pillars along the $y$-axis increases, leading to the splitting of these degenerate bands at the X-point (Figure 2). Each mode's sensitivity to changes in $q$ is closely related to its spatial displacement profile. As illustrated in Figure 3, modes in which all pillars along the $y$-axis oscillate in-phase, i.e., modes 1, 3, 6, and 7, indicated by solid curves, exhibit minimal frequency shifts and slope variations with changes in $q$. The in-phase oscillation of pillars across the lattice produces consistent coupling between neighboring pillars that remains relatively unaffected by lateral spacing changes. In contrast, modes 2, 4, 5, and 8, indicated by dashed curves, undergo significant frequency shifts and slope changes as $q$ varies. In these modes, neighboring pillars along the $y$-axis move in opposite directions, making their interaction highly sensitive to lateral spacing. Unlike the rectangular PnC, the band structure of the triangular PnC exhibits greater robustness against variations in $q$. While slope changes due to variations in $q$ also occur in the triangular PnC, degeneracy is preserved at the X-point for all $q$ values shown in Figure S1. This behavior is attributed to the higher inherent symmetry of the triangular lattice compared to the rectangular lattice.

The splitting of degenerate modes at specific $q$ values in the rectangular PnC creates extremely flat bands at the band edges. Compared to the degenerate modes in the triangular PnC, these non-degenerate flat bands yield lower SAW group velocities and stronger resonance effects, as shown in Figure 4, 5, S4 and S5. Moreover, they exhibit relatively larger frequency shifts by altering the lateral spacing of the pillars. Therefore, when rectangular PnCs are precisely designed and fabricated, a PnC mode can be aligned with the desired frequency, i.e., the frequency of the SAW generated by the input IDT. As a result, the SAW propagates through the PnC with an effectively lower velocity, resulting in an increased interaction between the SAW and the magnetostrictive material.

In our study, we achieved a normalized overall sensitivity of approximately 15.7 °/mT per 1 μm length of the delay line, which significantly exceeds previously reported values. For example, Kittmann et al. reported a sensitivity of 450 °/mT for a Love-wave SAW magnetic field sensor.[21] With a 3.8 mm long delay line, this corresponds to a normalized sensitivity of 0.12 °/mT·μm. Later, Schell et al. achieved a maximum sensitivity of 2000 °/mT by controlling the magnetic anisotropy of FeCoSiB during the deposition process [74] and developing an exchange bias stack of ferromagnetic and antiferromagnetic layers.[75] With the same delay line length, their normalized sensitivity is 0.53 °/mT·μm. Schmalz et al. reported a sensitivity of 1250 °/mT using a 1.8 mm delay line, resulting in a normalized sensitivity of 0.69 °/mT·μm.[76] These comparisons demonstrate that our sensor's sensitivity significantly surpasses existing SAW magnetic field sensors, highlighting the effectiveness of our design.

## 6. Conclusions

We demonstrate that the sensitivity of a SAW magnetic field sensor can be substantially enhanced by adding PnC structures. The presented design offers an eight-fold improvement in overall sensitivity compared to a PnC composed of FeCoSiB pillars and nearly two orders of magnitude improvement over continuous delay line structures. These improvements arise from embedding the pillars within the $SiO_2$ guiding layer, so that the PnC thickness is equal to that of the guiding layer, rather than being limited by the relatively thin FeCoSiB layer. As predicted in an earlier study[67], thicker PnCs produce comparatively stronger resonances. Additionally, in the present design, we avoid patterning the FeCoSiB layer in order to increase the interaction volume between the SAW and the magnetostrictive material.

In conclusion, our study highlights the potential of PnC-based designs to advance SAW sensor technology, paving the way for next-generation magnetic field sensors with unprecedented sensitivity and limit of detection. This advancement allows for more precise magnetic field detection across applications from industrial monitoring to medical diagnostics. While the simulation results demonstrate a significant sensitivity enhancement through the integration of PnCs, experimental validation is essential to confirm these findings. Future work will focus on fabricating the proposed sensor design and conducting experimental measurements to verify the predicted sensitivity improvements.

## 7. Material Parameters

In our model, we used Silicon as the substrate with $E$ = 169 GPa, $\upsilon$ = 0.28 and $\rho$ = 2329 kg/m$^3$ [69,77] and $Al_{0.77}Sc_{0.23}N$ as the piezoelectric material, whose material parameters are as follows:[78]

$$\mathbf{C}_{\text{AlScN}} = \begin{pmatrix} 335 & 154 & 119 & 0 & 0 & 0 \\ 154 & 335 & 119 & 0 & 0 & 0 \\ 119 & 119 & 270 & 0 & 0 & 0 \\ 0 & 0 & 0 & 109 & 0 & 0 \\ 0 & 0 & 0 & 0 & 109 & 0 \\ 0 & 0 & 0 & 0 & 0 & 90 \end{pmatrix} \text{GPa},$$

$$\boldsymbol{e}_{\text{AlScN}} = \begin{pmatrix} 0 & 0 & 0 & 0 & -0.25 & 0 \\ 0 & 0 & 0 & -0.25 & 0 & 0 \\ -0.6 & -0.6 & 2.13 & 0 & 0 & 0 \end{pmatrix} \text{C/m}^2,$$

$$\rho_{\text{AlScN}} = 3278 \text{ kg/m}^3,$$

$$\boldsymbol{\varepsilon}_{\text{r}_{\text{AlScN}}} = \begin{pmatrix} 9.5 & 0 & 0 \\ 0 & 9.5 & 0 \\ 0 & 0 & 11.5 \end{pmatrix},$$

$$\boldsymbol{\mu}_{\text{r}_{\text{AlScN}}} = \begin{pmatrix} 1 & 0 & 0 \\ 0 & 1 & 0 \\ 0 & 0 & 1 \end{pmatrix},$$

where $\mathbf{C}_{\text{AlScN}}$, $\boldsymbol{e}_{\text{AlScN}}$, and $\rho_{\text{AlScN}}$ represent the stiffness tensor, the piezoelectric coupling tensor, and the mass density of $Al_{0.77}Sc_{0.23}N$, respectively, while $\boldsymbol{\varepsilon}_{\text{r}_{\text{AlScN}}}$ and $\boldsymbol{\mu}_{\text{r}_{\text{AlScN}}}$ are the relative electrical permittivity and relative magnetic permeability of $Al_{0.77}Sc_{0.23}N$.

As the magnetostrictive material, we used FeCoSiB with a mass density of $\rho$ = 7250 kg/m$^3$, a Young's modulus of $E_0$ = 150 GPa and a Poisson's ratio of $\upsilon$ = 0.3 at fixed magnetization.[68] For the magnetoelastic simulations, we set the saturation flux density to $\mu_0 M_s$ = 1.5 T and the saturation magnetostriction to $\lambda_s$ = 35 ppm.[20,79] The first-order anisotropy constant was set to $K$ = 1400 J·m$^{-3}$.[80] The magnetization curve for the FeCoSiB layer, obtained using the aforementioned parameters, is presented in Figure S8. The stiffness tensors corresponding to the magnetic field values of $H^+$ and $H^+$ are given as:

$$\mathbf{C}^+ = \begin{pmatrix} 183.30 & 105.16 & 86.54 & 0 & 0 & -0.94 \\ 105.16 & 183.30 & 86.54 & 0 & 0 & 0.94 \\ 86.54 & 86.54 & 201.92 & 0 & 0 & 0 \\ 0 & 0 & 0 & 57.69 & 0 & 0 \\ 0 & 0 & 0 & 0 & 57.69 & 0 \\ -0.94 & 0.94 & 0 & 0 & 0 & 57.64 \end{pmatrix} \text{GPa},$$

$$\mathbf{C}^- = \begin{pmatrix} 184.81 & 103.66 & 86.54 & 0 & 0 & -1.19 \\ 103.66 & 184.81 & 86.54 & 0 & 0 & 1.19 \\ 86.54 & 86.54 & 201.92 & 0 & 0 & 0 \\ 0 & 0 & 0 & 57.69 & 0 & 0 \\ 0 & 0 & 0 & 0 & 57.69 & 0 \\ -1.19 & 1.19 & 0 & 0 & 0 & 57.61 \end{pmatrix} \text{GPa}.$$

**Acknowledgements**

This work was funded by the German Research Foundation (Deutsche Forschungsgemeinschaft) via the collaborative research center CRC 1261 "Magnetoelectric Sensors: From Composite Materials to Biomagnetic Diagnostics".

**References**

[1] P. Ripka, K. Závěta, *Handbook of Magnetic Materials* **2009**, *18*, 347.
[2] C. P. O. Treutler, *Sens Actuators A Phys* **2001**, *91*, 2.
[3] S. S. P. Parkin, C. Kaiser, A. Panchula, P. M. Rice, B. Hughes, M. Samant, S. H. Yang, *Nature Materials 2004 3:12* **2004**, *3*, 862.
[4] J. Zhai, Z. Xing, S. Dong, J. Li, D. Viehland, *Appl Phys Lett* **2006**, *88*, 62510.
[5] I. R. McFadyen, E. E. Fullerton, M. J. Carey, *MRS Bull* **2006**, *31*, 379.
[6] M. Díaz-Michelena, *Sensors 2009, Vol. 9, Pages 2271-2288* **2009**, *9*, 2271.
[7] D. Karnaushenko, D. D. Karnaushenko, D. Makarov, S. Baunack, R. Schäfer, O. G. Schmidt, *Advanced Materials* **2015**, *27*, 6582.
[8] G. Lin, D. Makarov, O. G. Schmidt, *Lab Chip* **2017**, *17*, 1884.
[9] S. Zuo, J. Schmalz, M. Ö. Özden, M. Gerken, J. Su, F. Niekiel, F. Lofink, K. Nazarpour, H. Heidari, *IEEE Trans Biomed Circuits Syst* **2020**, *14*, 971.
[10] D. Murzin, D. J. Mapps, K. Levada, V. Belyaev, A. Omelyanchik, L. Panina, V. Rodionova, *Sensors 2020, Vol. 20, Page 1569* **2020**, *20*, 1569.
[11] E. Elzenheimer, C. Bald, E. Engelhardt, J. Hoffmann, P. Hayes, J. Arbustini, A. Bahr, E. Quandt, M. Höft, G. Schmidt, *Sensors 2022, Vol. 22, Page 1018* **2022**, *22*, 1018.
[12] Hanna, S. M., *ITUFF* **1987**, *34*, 191.
[13] V. Polewczyk, K. Dumesnil, D. Lacour, M. Moutaouekkil, H. Mjahed, N. Tiercelin, S. Petit Watelot, H. Mishra, Y. Dusch, S. Hage-Ali, O. Elmazria, F. Montaigne, A. Talbi, O. Bou Matar, M. Hehn, *Phys Rev Appl* **2017**, *8*, 024001.
[14] W. Wang, Y. Jia, X. Xue, Y. Liang, Z. Du, *AIP Adv* **2018**, *8*, DOI 10.1063/1.5012579/990884.
[15] X. Liu, B. Tong, J. Ou-Yang, X. Yang, S. Chen, Y. Zhang, B. Zhu, *Appl Phys Lett* **2018**, *113*, DOI 10.1063/1.5044478/37057.
[16] E. W. Lee, *Reports on Progress in Physics* **1955**, *18*, 184.
[17] J. D. Livingston, *physica status solidi (a)* **1982**, *70*, 591.
[18] R. Kellogg, A. Flatau, *http://dx.doi.org/10.1177/1045389X07077854* **2007**, *19*, 583.
[19] S. Datta, J. Atulasimha, C. Mudivarthi, A. B. Flatau, *J Magn Magn Mater* **2010**, *322*, 2135.
[20] B. Spetzler, E. V. Golubeva, R. M. Friedrich, S. Zabel, C. Kirchhof, D. Meyners, J. McCord, F. Faupel, *Sensors 2021, Vol. 21, Page 2022* **2021**, *21*, 2022.
[21] A. Kittmann, P. Durdaut, S. Zabel, J. Reermann, J. Schmalz, B. Spetzler, D. Meyners, N. X. Sun, J. McCord, M. Gerken, G. Schmidt, M. Höft, R. Knöchel, F. Faupel, E. Quandt, *Scientific Reports* **2018**, *8*, 1.
[22] A. Kittmann, C. Müller, P. Durdaut, L. Thormählen, V. Schell, F. Niekiel, F. Lofink, D. Meyners, R. Knöchel, M. Höft, J. McCord, E. Quandt, *Sens Actuators A Phys* **2020**, *311*, 111998.
[23] J. Su, F. Niekiel, S. Fichtner, C. Kirchhof, D. Meyners, E. Quandt, B. Wagner, F. Lofink, *Journal of Micromechanics and Microengineering* **2020**, *30*, 075009.

[24] P. Durdaut, C. Müller, A. Kittmann, V. Schell, A. Bahr, E. Quandt, R. Knöchel, M. Höft, J. McCord, *Sensors 2021, Vol. 21, Page 5631* **2021**, *21*, 5631.
[25] J. M. Meyer, V. Schell, J. Su, S. Fichtner, E. Yarar, F. Niekiel, T. Giese, A. Kittmann, L. Thormählen, V. Lebedev, S. Moench, A. Žukauskaitė, E. Quandt, F. Lofink, *Sensors 2021, Vol. 21, Page 8166* **2021**, *21*, 8166.
[26] H. Greve, E. Woltermann, R. Jahns, S. Marauska, B. Wagner, R. Knöchel, M. Wuttig, E. Quandt, *Appl Phys Lett* **2010**, *97*, DOI 10.1063/1.3497277/122342.
[27] S. Zabel, J. Reermann, S. Fichtner, C. Kirchhof, E. Quandt, B. Wagner, G. Schmidt, F. Faupel, *Appl Phys Lett* **2016**, *108*, DOI 10.1063/1.4952735/30497.
[28] S. Salzer, V. Röbisch, M. Klug, P. Durdaut, J. McCord, D. Meyners, J. Reermann, M. Höft, R. Knöchel, *IEEE Sens J* **2018**, *18*, 596.
[29] M. Yamaguchi, M. Naoe, H. Kogo, *IEEE Trans Magn* **1980**, *16*, 916.
[30] Y. Q. Fu, J. K. Luo, N. T. Nguyen, A. J. Walton, A. J. Flewitt, X. T. Zu, Y. Li, G. McHale, A. Matthews, E. Iborra, H. Du, W. I. Milne, *Prog Mater Sci* **2017**, *89*, 31.
[31] Y. Takagaki, P. V. Santos, E. Wiebicke, O. Brandt, H. P. Schönherr, K. H. Ploog, *Appl Phys Lett* **2002**, *81*, 2538.
[32] G. Piazza, V. Felmetsger, P. Muralt, R. H. Olsson, R. Ruby, *MRS Bull* **2012**, *37*, 1051.
[33] U. C. Kaletta, P. V. Santos, D. Wolansky, A. Scheit, M. Fraschke, C. Wipf, P. Zaumseil, C. Wenger, *Semicond Sci Technol* **2013**, *28*, 065013.
[34] T. Aubert, J. Bardong, O. Legrani, O. Elmazria, M. Badreddine Assouar, G. Bruckner, A. Talbi, *J Appl Phys* **2013**, *114*.
[35] R. M. R. Pinto, V. Gund, R. A. Dias, K. K. Nagaraja, K. B. Vinayakumar, *Journal of Microelectromechanical Systems* **2022**, *31*, 500.
[36] S. Fichtner, N. Wolff, G. Krishnamurthy, A. Petraru, S. Bohse, F. Lofink, S. Chemnitz, H. Kohlstedt, L. Kienle, B. Wagner, *J Appl Phys* **2017**, *122*.
[37] J. Casamento, C. S. Chang, Y. T. Shao, J. Wright, D. A. Muller, H. Xing Grace, D. Jena, *Appl Phys Lett* **2020**, *117*.
[38] L. Dreher, M. Weiler, M. Pernpeintner, H. Huebl, R. Gross, M. S. Brandt, S. T. B. Goennenwein, *Phys Rev B Condens Matter Mater Phys* **2012**, *86*.
[39] X. Li, D. Labanowski, S. Salahuddin, C. S. Lynch, *J Appl Phys* **2017**, *122*, 43904.
[40] Y. Nozaki, S. Yanagisawa, *Electrical Engineering in Japan* **2018**, *204*, 3.
[41] D. A. Bas, P. J. Shah, M. E. McConney, M. R. Page, *J Appl Phys* **2019**, *126*, 114501.
[42] R. Sasaki, Y. Nii, Y. Iguchi, Y. Onose, *Phys Rev B* **2017**, *95*.
[43] R. Verba, V. Tiberkevich, A. Slavin, *Phys Rev Appl* **2019**, *12*.
[44] M. Xu, K. Yamamoto, J. Puebla, K. Baumgaertl, B. Rana, K. Miura, H. Takahashi, D. Grundler, S. Maekawa, Y. Otani, *Sci Adv* **2020**, *6*.
[45] P. J. Shah, D. A. Bas, I. Lisenkov, A. Matyushov, N. X. Sun, M. R. Page, *Sci Adv* **2020**, *6*.
[46] C. Chen, L. Han, P. Liu, Y. Zhang, S. Liang, Y. Zhou, W. Zhu, S. Fu, F. Pan, C. Song, *Advanced Materials* **2023**, *35*, 2302454.
[47] E. Yablonovitch, *JOSA B* **1993**, *10*, 283.
[48] T. F. Krauss, R. M. De La Rue, S. Brand, *Nature* **1996**, *383*, 699.
[49] S. Noda, K. Tomoda, N. Yamamoto, A. Chutinan, *Science* **2000**, *289*, 604.
[50] Y. Akahane, T. Asano, B. S. Song, S. Noda, *Nature* **2003**, *425*, 944.
[51] P. Lalanne, C. Sauvan, J. P. Hugonin, *Laser Photon Rev* **2008**, *2*, 514.
[52] S. Hu, S. M. Weiss, *ACS Photonics* **2016**, *3*, 1647.
[53] J. D. Joannopoulos, M. Soljačić, E. Ippen, M. Ibanescu, S. Fan, S. G. Johnson, *JOSA B,* **2002**, *19*, 2052.
[54] M. Dorseth, S. Bourouaine, J. C. Perez, T. Xie, Z. Wu, T. J. Foutz, N. K. Walia, K. Seki, T. F. Krauss, *J Phys D Appl Phys* **2007**, *40*, 2666.
[55] T. Baba, *Nature Photonics* **2008**, *2*, 465.

[56] D. Threm, Y. Nazirizadeh, M. Gerken, *J Biophotonics* **2012**, *5*, 601.
[57] R. Gao, Y. Jiang, S. Abdelaziz, *Optics Letters* **2013**, *38*, 1539.
[58] Y. N. Zhang, Y. Zhao, T. Zhou, Q. Wu, *Lab Chip* **2017**, *18*, 57.
[59] A. Khelif, A. Adibi, *Phononic Crystals: Fundamentals and Applications* **2015**, 1.
[60] M. Krawczyk, D. Grundler, *Journal of Physics: Condensed Matter* **2014**, *26*, 123202.
[61] S. Tamura, D. C. Hurley, J. P. Wolfe, *Phys Rev B* **1988**, *38*, 1427.
[62] V. Laude, M. Wilm, S. Benchabane, A. Khelif, *Phys Rev E Stat Nonlin Soft Matter Phys* **2005**, *71*, 036607.
[63] S. Benchabane, A. Khelif, J. Y. Rauch, L. Robert, V. Laude, *Phys Rev E Stat Nonlin Soft Matter Phys* **2006**, *73*, 065601.
[64] A. Khelif, B. Aoubiza, S. Mohammadi, A. Adibi, V. Laude, *Phys Rev E Stat Nonlin Soft Matter Phys* **2006**, *74*, 046610.
[65] M. Badreddine Assouar, M. Oudich, *Applied Physics Letters* **2011**, *99*, 123505.
[66] M. Oudich, N. J. Gerard, Y. Deng, Y. Jing, M. Oudich, N. J. Gerard, Y. Deng, Y. Jing, M. O. Institut, J. Lamour, *Advanced Functional Materials* **2023**, *33*, 2206309.
[67] M. Samadi, J. Schmalz, J. M. Meyer, F. Lofink, M. Gerken, *Micromachines* **2023**, *14*, 2130.
[68] C. Cordier, C. Dolabdjian, *IEEE Sensors Journal* **2023**, *23*, 2014.
[69] J. Schmalz, E. Spetzler, J. McCord, M. Gerken, *Sensors* **2023**, *23*, 5012.
[70] J. Yang, *An Introduction to the Theory of Piezoelectricity* **2018**, *9*.
[71] IEEE Standard on Magnetostrictive Materials: Piezomagnetic Nomenclature, *IEEE Transactions on Sonics and Ultrasonics* **1973**, *20*, 67.
[72] COMSOL Multiphysics® v. 6.2. COMSOL AB, Stockholm, Sweden.
[73] A. Khelif, Y. Achaoui, S. Benchabane, V. Laude, B. Aoubiza, *Physical Review B - Condensed Matter and Materials Physics* **2010**, *81*, 214303.
[74] V. Schell, C. Müller, P. Durdaut, A. Kittmann, L. Thormählen, F. Lofink, D. Meyners, M. Höft, J. McCord, E. Quandt, *Appl Phys Lett* **2020**, 116, 7.
[75] V. Schell, E. Spetzler, N. Wolff, L. Bumke, L. Kienle, J. McCord, E. Quandt, D. Meyners, *Scientific Reports* **2023**, 13, 1, 8446.
[76] J. Schmalz, A. Kittmann, P. Durdaut, B. Spetzler, F. Faupel, M. Höft, E. Quandt, M. Gerken, *Sensors* **2020**, 20, 12, 3421.
[77] J. L. Gugat, M. C. Krantz, M. Gerken, *IEEE Transactions on Magnetics* **2013**, *49*, 5287.
[78] M. A. Caro, S. Zhang, T. Riekkinen, M. Ylilammi, M. A. Moram, O. Lopez-Acevedo, J. Molarius, T. Laurila, *Journal of Physics: Condensed Matter* **2015**, *27*, 245901.
[79] A. Ludwig, E. Quandt, *IEEE Transactions on Magnetics* **2002**, *38*, 2829.
[80] B. Spetzler, E. V. Golubeva, C. Müller, J. McCord, F. Faupel, *Sensors* **2019**, *19*, 4769.

Supplementary Information

# Modeling of High-Sensitivity SAW Magnetic Field Sensors with Au-$SiO_2$ Phononic Crystals

*Mohsen Samadi*[1,*], *Jana Marie Meyer*[2], *Elizaveta Spetzler*[3], *Benjamin Spetzler*[4], *Jeffrey McCord*[3,5], *Fabian Lofink*[2,5,6], *and Martina Gerken*[1,5,*]

[1]Integrated Systems and Photonics, Department of Electrical and Information Engineering, Kiel University, Kaiserstraße 2, 24143 Kiel, Germany

[2]Fraunhofer Institute for Silicon Technology ISIT, 25524 Itzehoe, Germany

[3]Nanoscale Magnetic Materials - Magnetic Domains, Department of Materials Science, Kiel University, Kaiserstraße 2, 24143 Kiel, Germany

[4]Energy Materials and Devices, Department of Materials Science, Kiel University, Kaiserstraße 2, 24143 Kiel, Germany

[5]Kiel Nano, Surface and Interface Science (KiNSIS), Kiel University, Kaiserstraße 2, 24143 Kiel, Germany

[6]Microsystem Materials, Department of Materials Science, Kiel University, Kaiserstraße 2, 24143 Kiel, Germany

*E-Mail: mosa@tf.uni-kiel.de, mge@tf.uni-kiel.de

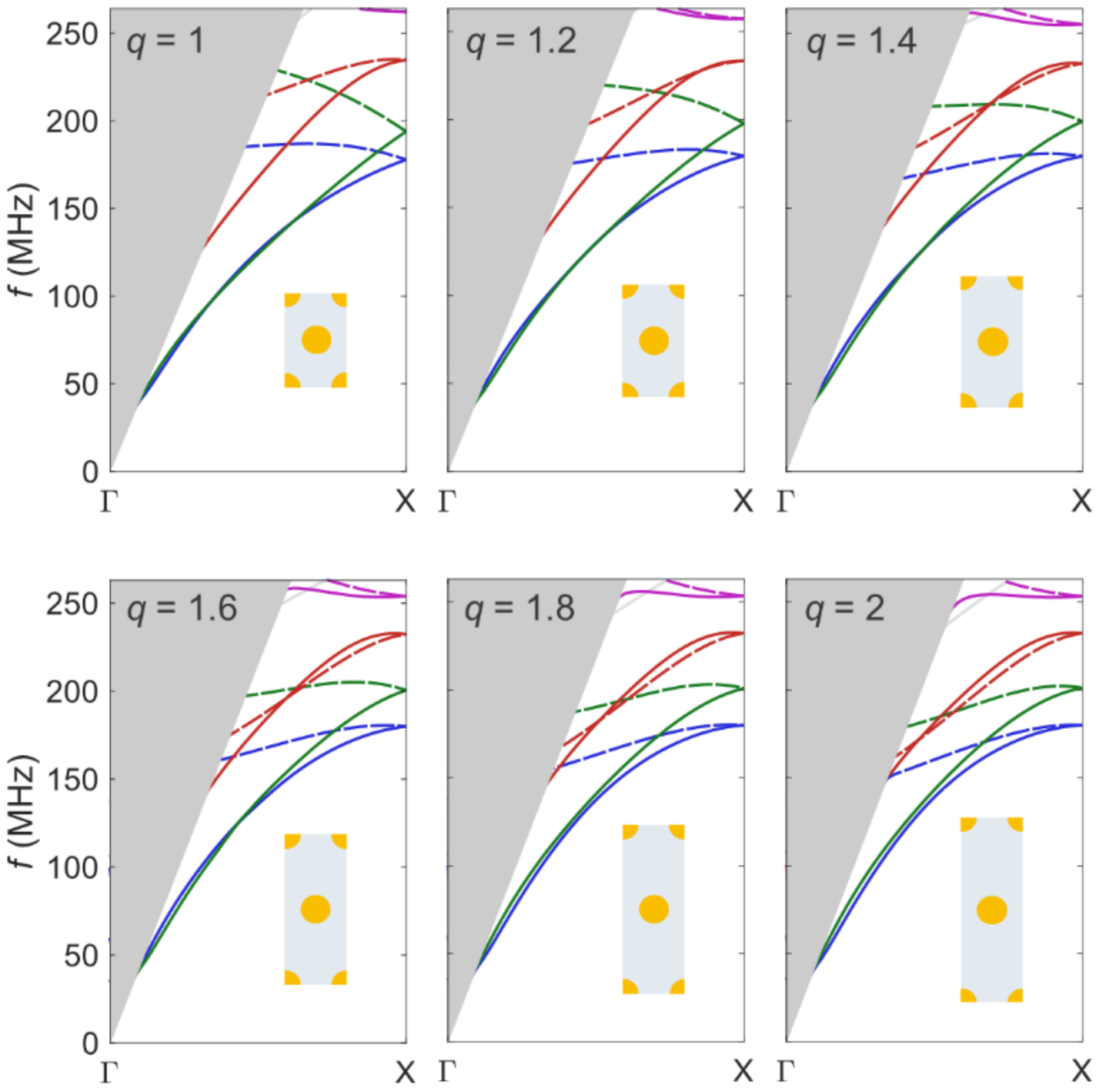

**Figure S1.** Band structures of the triangular PnC, as illustrated in Figure 1(d), with $a$ = 6 µm, $r_p$ = 1.5 µm, and different values of the tuning coefficient: $q$=1, $q$=1.2, $q$=1.4, $q$=1.6, $q$=1.8 and $q$=2. The lattice constant along the $y$-axis is expressed as $b = \sqrt{3}aq$ for the triangular PnC. All diagrams were calculated along the Γ-X direction of the first Brillouin zone. The grey-shaded area outlines the sound cone.

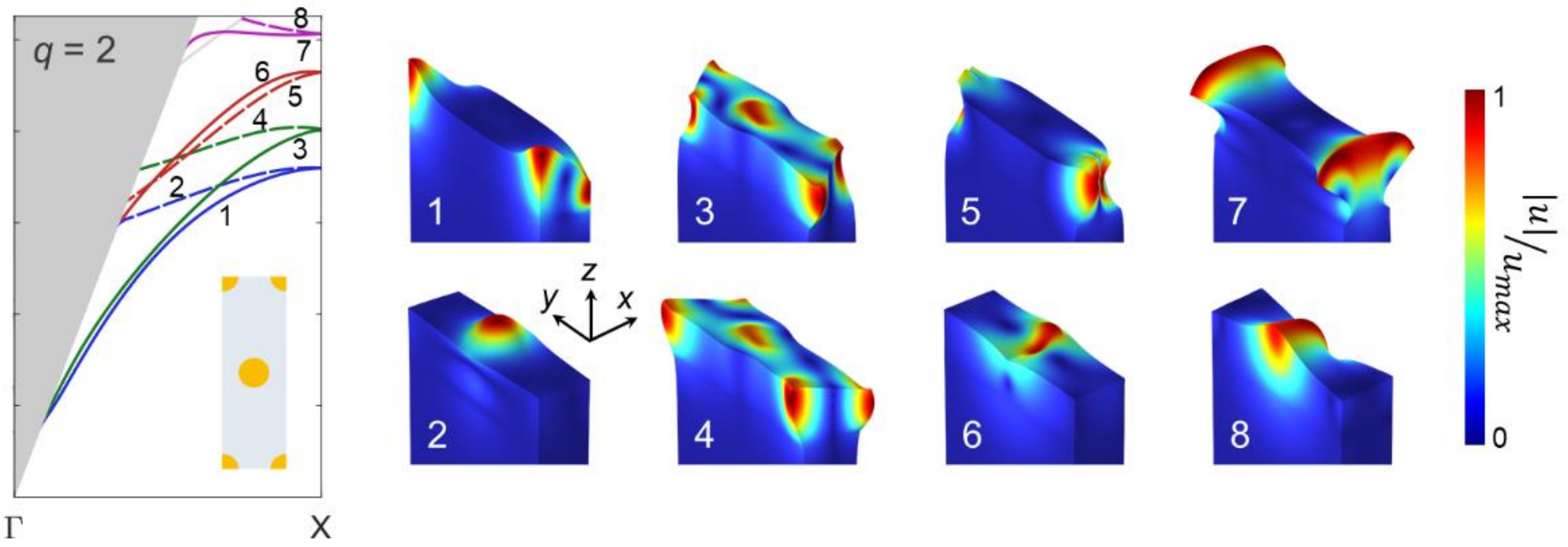

**Figure S2.** Band structure of the triangular PnC with $a$ = 6 µm, $r_p$ = 1.5 µm, and $q$=2, calculated along the Γ-X direction of the first Brillouin zone. Distinct surface acoustic modes are labeled by numbers, with corresponding normalized modal displacements ($|u|/u_{max}$) displayed within a single supercell at a wavevector $k$ near the edge of the first Brillouin zone. Lower-order modes exhibit almost uniaxial oscillations: modes 1 and 2 (blue) along the $z$-axis, modes 3 and 4 (green) along the $y$-axis, and modes 5 and 6 (red) along the $x$-axis. In contrast, higher-order modes 7 and 8 (magenta) exhibit coupled longitudinal and vertical oscillations in the $x$-$z$ plane.

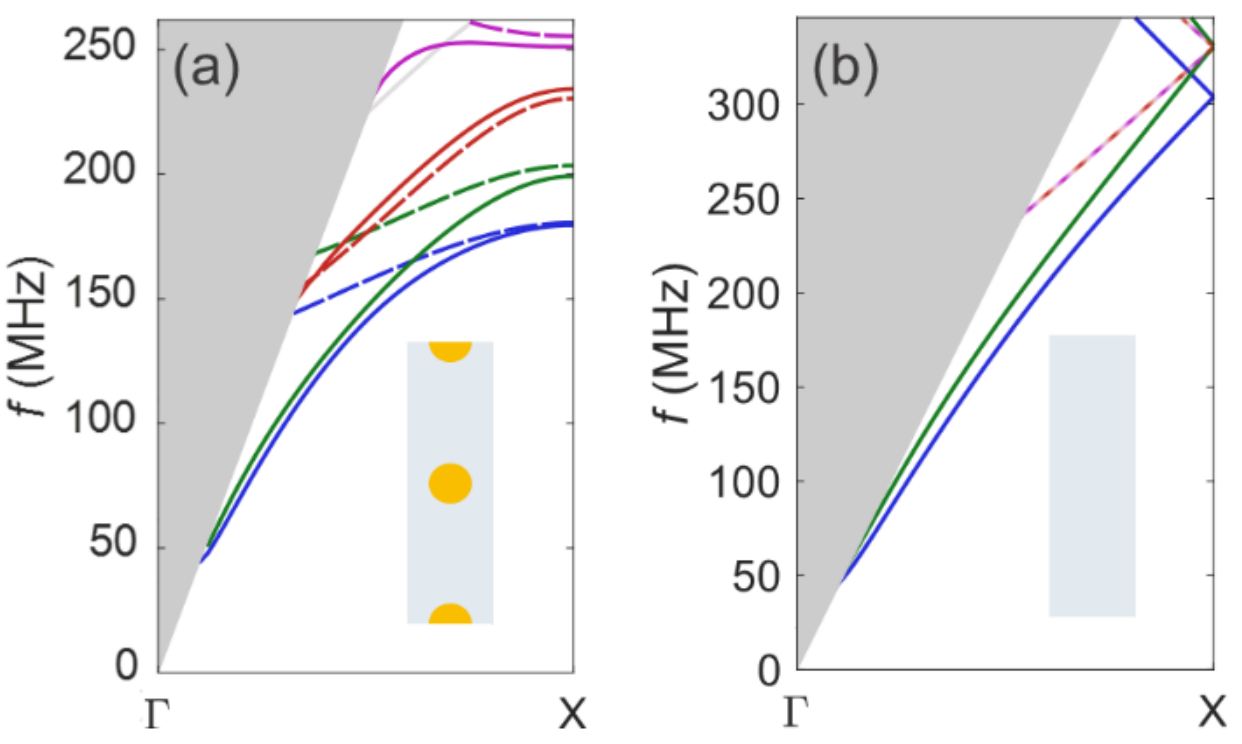

**Figure S3.** Band structures of (a) rectangular PnC with $a$ = 6 μm, $r_p$ = 1.5 μm, and $q$=2 and (b) continuous guiding layer with similar dimensions, calculated along the Γ-X direction of the first Brillouin zone. The insets show the top views of the rectangular supercells used for the calculations. Blue, green and red lines correspond to surface acoustic modes with uniaxial oscillations along the *z*-, *y*- and *x*-axes, respectively, while magenta lines represent Rayleigh modes with both longitudinal and vertical surface displacements in the *x*-*z* plane. The integration of the PnC flattens all bands and lifts the degeneracy of hybrid modes (red-magenta lines in (b)), resulting in their splitting into distinct longitudinal (red) and Rayleigh (magenta) modes.

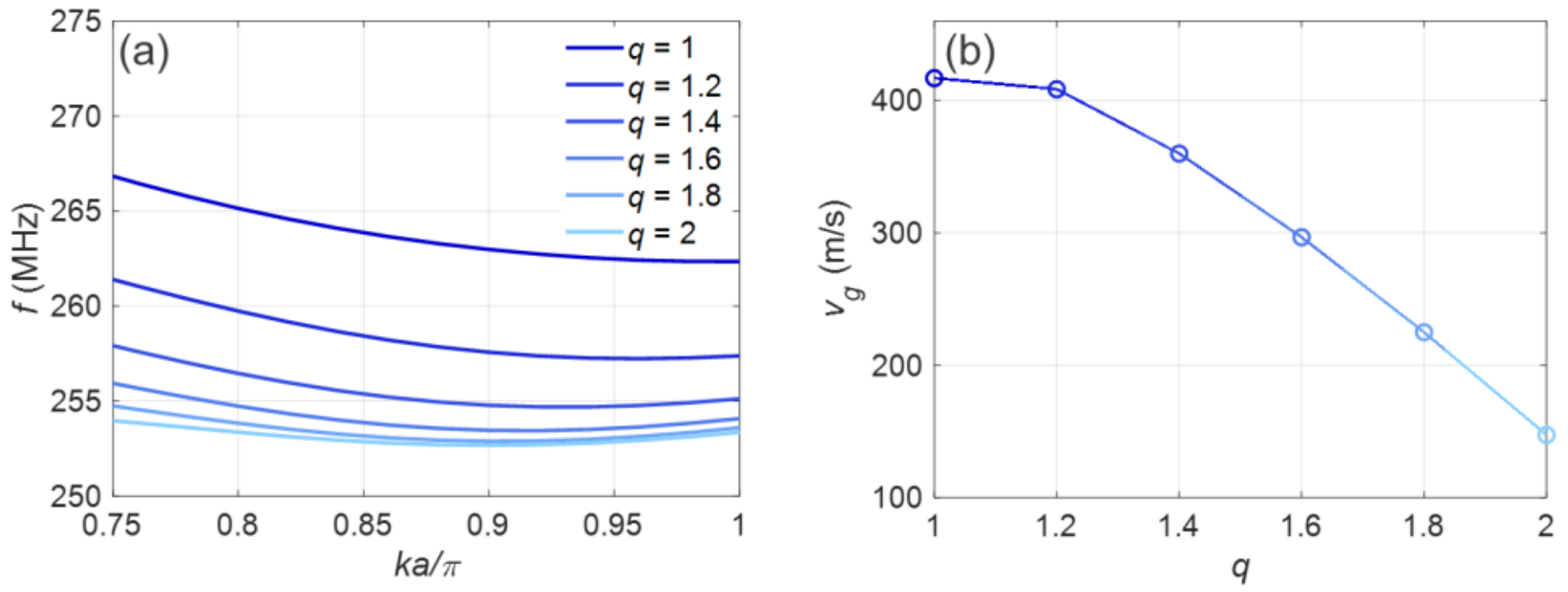

**Figure S4.** (a) Dispersion curves of mode 7 near the edge of the first Brillouin zone for the triangular PnC at different values of the tuning coefficients: $q$=1, $q$=1.2, $q$=1.4, $q$=1.6, $q$=1.8 and $q$=2. (b) Group velocity variations with $q$ for the triangular PnC, calculated from the dispersion curves at $k = 0.8\pi/a$.

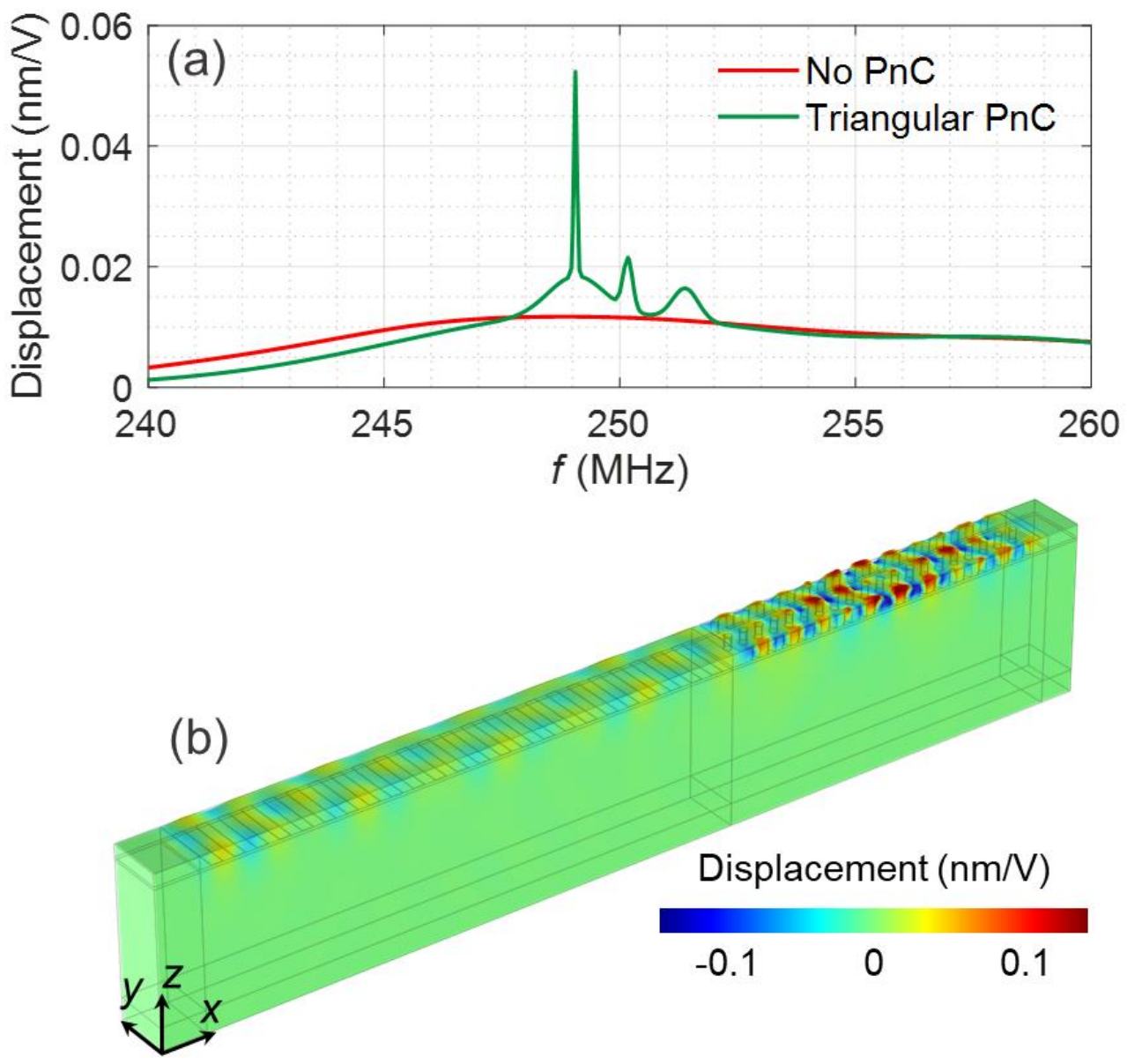

**Figure S5.** (a) Average displacement spectra across the top surface of the delay line with the triangular PnC with $a$ = 6 µm, $r_p$ = 1.5 µm, and $q$=2 (green curve). The average displacement spectra for a continuous delay line without PnC is also plotted for comparison (red curve). (b) The spatial displacement along the triangular PnC at its resonance frequency. The displacement values are normalized to the amplitude of the AC electric potential applied to the IDT. The PnC has 20 periodicities along the $x$-axis ($n$ = 20) and the IDT comprises 12 pairs of Au split-finger structures with a periodicity of 16 µm, a finger width of 4 µm, and a thickness of 150 nm.

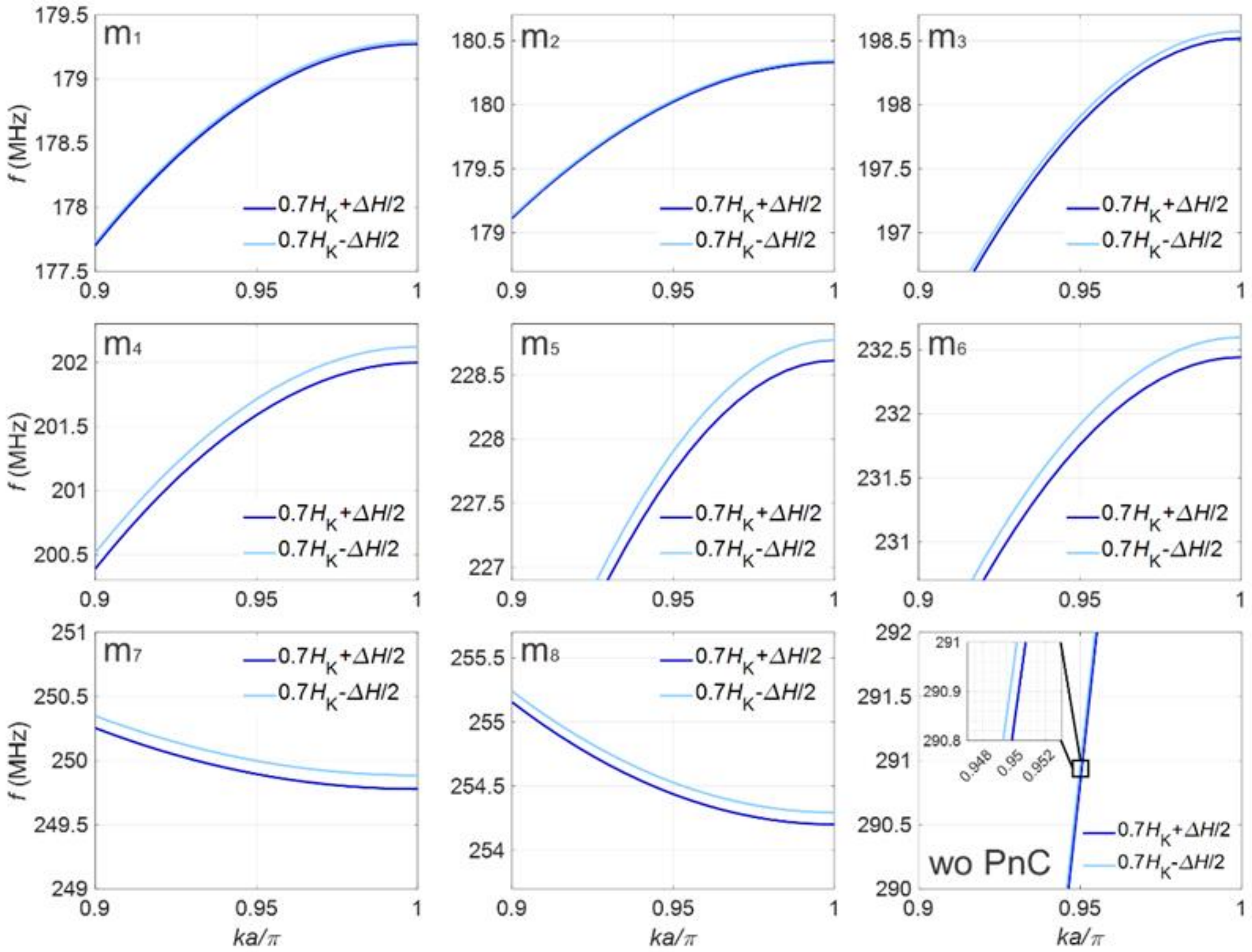

**Figure S6.** Dispersion curves of the rectangular PnC modes $m_1$ to $m_8$ displayed near the edge of the first Brillouin zone, at $H^+ = 0.7H_K + \Delta H/2$ and $H^- = 0.7H_K - \Delta H/2$, with $\Delta H = 100$ μT. The dispersion curves of the lowest-order mode of a continuous delay line without PnCs are also provided for comparison.

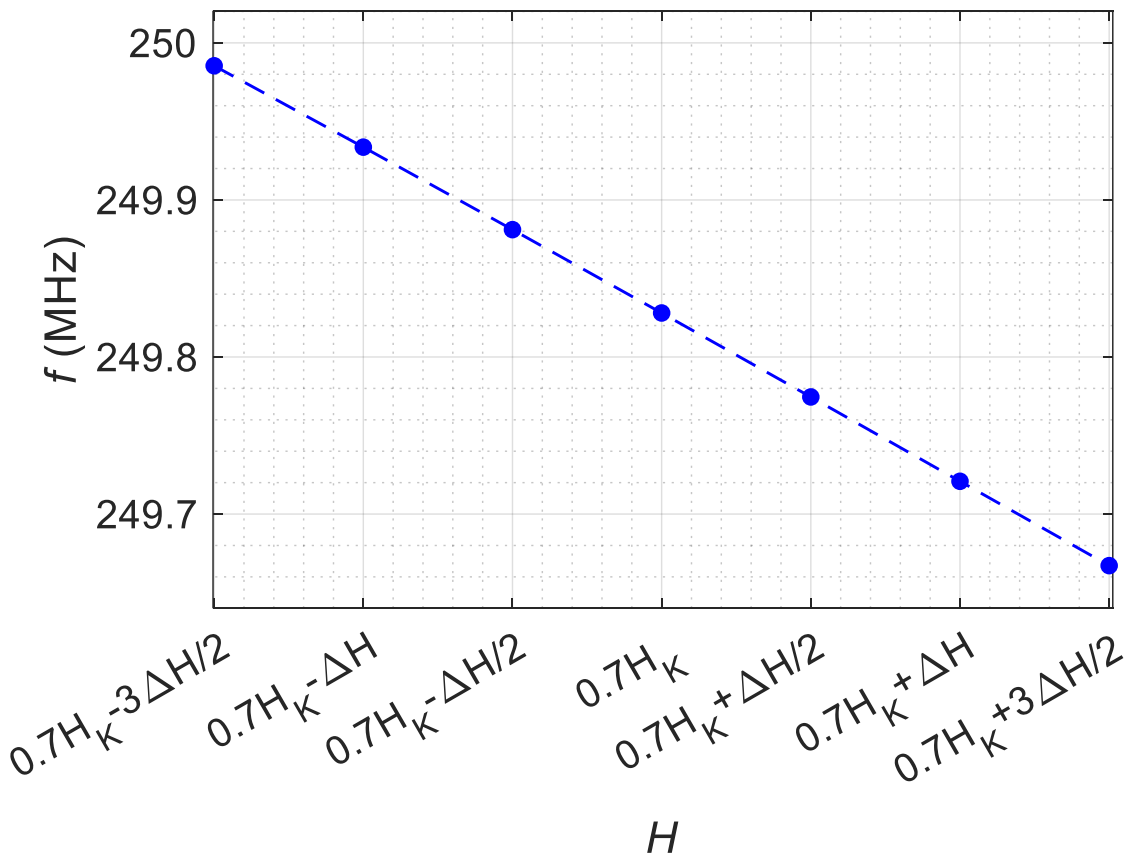

**Figure S7.** Resonance frequency of the PnC mode $m_7$ as a function of the applied magnetic field, expressed in units of $\Delta H = 100$ μT relative to the bias magnetic field $H_0 = 0.7H_K$.

**Table S1.** Magneto-structural sensitivity of the rectangular PnC modes $m_1$ to $m_8$ normalized to their respective resonance frequencies, calculated for different PnC thicknesses $h$ = 2.5, 3, 3.5, 4, 4.5 and 5 μm.

| | $h$ (μm) | $m_1$ | $m_2$ | $m_3$ | $m_4$ | $m_5$ | $m_6$ | $m_7$ | $m_8$ |
|---|---|---|---|---|---|---|---|---|---|
| $S_{\text{mag-str}}\left(\frac{\text{m/s}}{\text{mT.MHz}}\right)$ | 2.5 | 1.2 | 1.2 | 2.5 | 4.4 | 3.7 | 4.4 | 4.4 | 3.1 |
| | 3 | 1.2 | 1.2 | 2.5 | 3.7 | 3.7 | 3.7 | 9.8 | 3.1 |
| | 3.5 | 1.2 | 1.2 | 1.8 | 3.7 | 3.1 | 3.7 | 7.7 | 3.1 |
| | 4 | 1.2 | 1.2 | 1.8 | 3.7 | 3.1 | 3.1 | 5.7 | 3.7 |
| | 4.5 | 1.2 | 1.2 | 1.8 | 3.1 | 2.5 | 3.1 | 6.4 | 3.7 |
| | 5 | 1.2 | 1.2 | 1.8 | 3.1 | 2.5 | 2.5 | 7 | 4.4 |

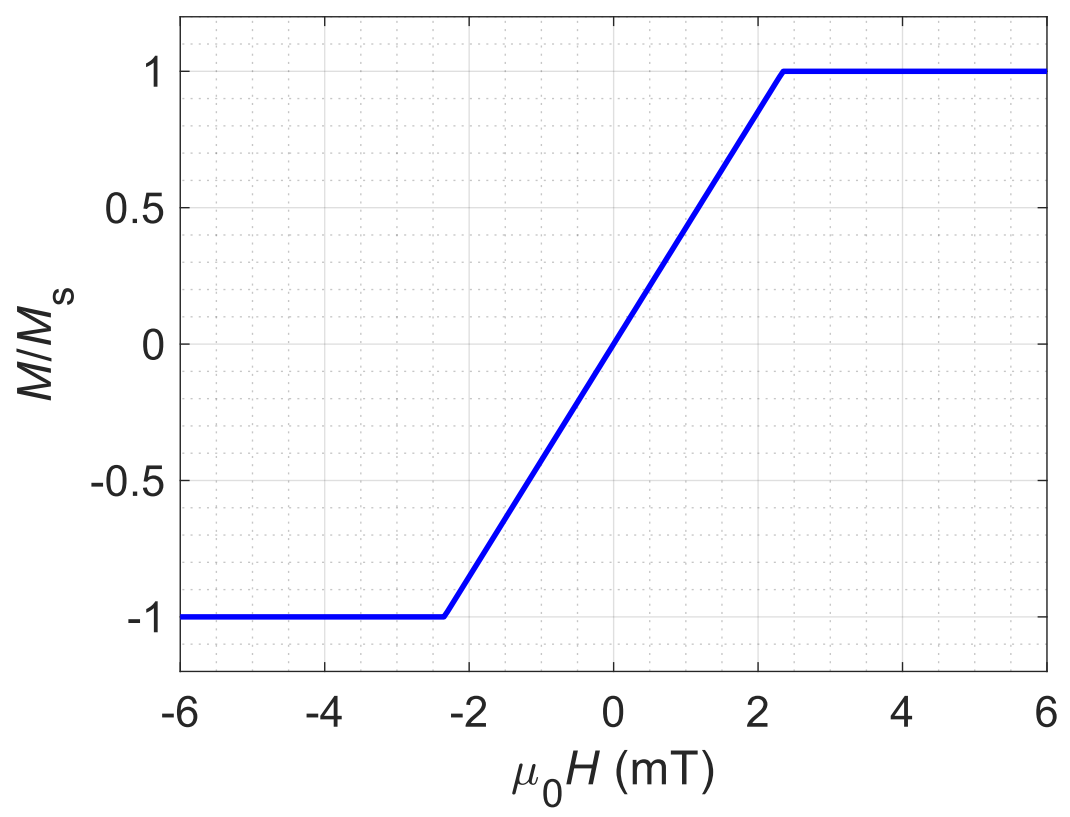

**Figure S8.** Magnetization curve of the FeCoSiB layer, normalized to the saturation magnetization $M_s$, as a function of the applied magnetic field $\mu_0 H$ along the hard axis. The saturation flux density and saturation magnetostriction were set to $\mu_0 M_s$ = 1.5 T and $\lambda_s$ = 35 ppm, respectively. The first-order anisotropy constant was set to $K$ = 1400 J·m$^{-3}$, yielding an effective anisotropy field of $\mu_0 H_K \simeq$ 2.35 mT.